\documentclass[twocolumn,amsmath,amssymb,pra,superscriptaddress]{revtex4-1}
\usepackage{graphicx}
\usepackage{dcolumn}
\usepackage{bm}
\usepackage{subfigure}
\usepackage{booktabs}
\usepackage{color}
\newcounter{one}
\def\bra#1{\mbox{\boldmath $#1$}^{\top}}
\def\ket#1{\mbox{\boldmath $#1$}}
\newcommand{\bracket}[1]{\left\langle #1 \right\rangle}

\newcommand{\cav}[2]{\backslash (#1,#2)}

\newcommand{\affA}{
Artificial Intelligence Research Center, 
National Institute of Advanced Industrial Science and Technology, 
2-3-26 Aomi, Koto-ku, Tokyo, Japan
}
\newcommand{\affB}{
Department of Mathematical and Computing Science, Tokyo Institute of Technology, 
W8-45, 2-12-1 Ookayma, Meguro-ku, Tokyo, Japan
}

\begin{document}

\title{\textbf{Comparative analysis on the selection of the number of clusters in community detection}}

\author{Tatsuro Kawamoto}
\affiliation{\affA}
\author{Yoshiyuki Kabashima}
\affiliation{\affB}
\date{\today}

\begin{abstract}
We conduct a comparative analysis on various estimates of the number of clusters in community detection. 
An exhaustive comparison requires testing of all possible combinations of frameworks, algorithms, and assessment criteria. 
In this paper we focus on the framework based on a stochastic block model, and investigate the performance of greedy algorithms, statistical inference, and spectral methods. 
For the assessment criteria, we consider modularity, map equation, Bethe free energy, prediction errors, and isolated eigenvalues. 
From the analysis, the tendency of overfit and underfit that the assessment criteria and algorithms have, becomes apparent. 
In addition, we propose that the alluvial diagram is a suitable tool to visualize statistical inference results and can be useful to determine the number of clusters. 
\end{abstract}
 
\maketitle

\section{Introduction}\label{Introduction}
Community detection is a coarse graining process for networks. 
Whereas the original dataset, given as a network, possesses information that is quite rich, it is often too microscopic to have its important structures interpreted. 
For better interpretability, a community detection algorithm summarizes (i.e., clustering or partitioning) the dataset by aggregating the vertices and edges of densely connected components. 
That is, the detailed relational information of similar vertices is discarded, while preserving an important macroscopic structure. 
A set of aggregated vertices is regarded as a cluster, or a community. 

A straightforward approach, or a framework, for community detection is to optimize an objective function that evaluates the quality of clustering. 
Another popular approach is based on statistical inference and considers a generative model of a network. It is often formulated using the so-called stochastic block model \cite{holland1983stochastic,WangWong87,KarrerNewman2011} as a generative model, which is a random graph with a modular structure. Therefore, the community structure can be inferred by fitting the network to the model. 
While these two approaches may seem very different, the former can sometimes be formulated as a limiting case (zero-temperature limit in physics terminology) of the latter, and in this paper, it is mainly explained in terms of the latter approach. 

In regards to these frameworks, a number of algorithms have been proposed, such as greedy algorithms \cite{Fortunato2010,Blondel2008,Rosvall2008}, spectral methods \cite{Luxburg2007,Newman2006PRE,Krzakala2013,SaadeBetheHessian}, and inference algorithms such as expectation-maximization (EM) algorithms \cite{daudin08,latouche12,Decelle2011a,Hayashi2016,KawamotoKabashimasbmBIX} and Monte Carlo methods \cite{Nowicki2001,PeixotoPRE2014MonteCarlo,NewmanReinert2016}, to name a few. 
In this paper, the following are considered: Louvain method \cite{Blondel2008} and Infomap \cite{Rosvall2008} for the greedy algorithms, the modularity matrix \cite{Newman2006politicalbooks,Newman2006PRE} and non-backtracking matrix \cite{Krzakala2013} for the matrices in the spectral methods, and the EM algorithm with belief propagation (BP) \cite{Decelle2011,KawamotoKabashimasbmBIX} for the inference algorithm. 

When community detection is performed, the number of clusters $q^{\ast}$ needs to be determined. In other words, the complexity of the model, i.e., the model selection needs to be specified. 
This process consists of determining the partition that describes the modular structure most efficiently, or to evaluate whether the obtained partition is statistically significant. 
Just as the quality of the $q$-way partition varies depending on the very definition of a cluster, the appropriateness of the number of clusters also varies depending on the principle followed. 
Meanwhile, it is difficult to decide on which principle to apply, given a dataset. 
Therefore, it is important to investigate the typical behavior and biases of each criterion. 
For example, some criteria may behave very differently from others in some cases, or some criteria may be more sensitive to the accuracy of a particular algorithm. 
Moreover, the dangers of underfit and overfit are often not symmetric. 
In the case of community detection, it is often safer to underfit than to overfit, because the former only implies a different level of coarse graining, while the latter implies the detection of fictitious small clusters. 

In this paper, a comparative analysis of various criteria that estimate $q^{\ast}$ is conducted. 
This analysis is distinct from other comparative analyses in the following sense. 
Whereas the performance of the community detection completely depends on the framework (or objective function), algorithm, and assessment criterion used, it is often the case that a specific combination of them is employed in a benchmark test. 
For example, Infomap, a greedy algorithm for the map equation \cite{Rosvall2008,Rosvall2011}, is a popular algorithm and frequently appears in benchmark tests. However, when the performance of a certain objective function or an assessment criterion is compared with the map equation, it is fair to use a common algorithm. 
For this reason, the performance of various assessment criteria using the same statistical inference algorithm is compared. 
In addition, the performance of the same assessment criterion on the same dataset using different algorithms is investigated. 
Therefore, when a criterion is ill-behaved, it can be argued whether it is due to the criterion itself, or the algorithm used. 

As can be observed below, sometimes, the validation curves of assessment criteria change very gradually. 
In such a case, it is not easy to determine the plausible value of $q^{\ast}$ from the assessment criteria, and a finer inspection of the partition actually obtained is required. 
For this purpose, a visualization technique, called the alluvial diagram \cite{Rosvall2010} is proposed as a suitable tool; not only because of the way the network is partitioned, but also because it allows the significance level to be evaluated from an inference algorithm.

The remainder of the paper is organized as follows. 
First, stochastic block models are defined in Sec.~\ref{Model} to set the basic framework. 
Second, the algorithms used to determine the cluster assignments and for the estimate of $q^{\ast}$ in Sec.~\ref{Algorithm} are introduced. 
Third, the assessment criteria of the number of clusters $q^{\ast}$ is explained in Sec.~\ref{ModelAssessmentCriteria}. 
Then, in Sec.~\ref{ComparativeAnalysis}, the results of the comparative analyses are shown. 
In Sec.~\ref{Visualization}, how the alluvial diagram helps determine the number of clusters is explained. 
Finally, Sec.~\ref{SummaryDiscussion} is devoted to the summary and discussion.


\section{Stochastic block models}\label{Model}
Community detection based on a stochastic block model is considered to be the basic framework. 
The sets of vertices and edges are denoted as $V$ and $E$, respectively. 
Their respective cardinalities are referred to as $N$ and $L$.

\subsection{Standard stochastic block model}
The most standard version of stochastic block models is constructed as follows. 
We first consider a set of vertices $V$ without edges. 
For each vertex, we randomly specify the cluster assignment $\sigma_{i} \in \{1,\dots, q\}$, where $i$ is the index of a vertex and the number of clusters $q$ is given as an input. 
The probability of the cluster size can also be specified. 
It is a prior distribution of the cluster assignments and is expressed by a multinomial distribution $\prod_{i} \gamma_{\sigma_{i}}$. 
Then, the edges are generated randomly according to the vertex pair's cluster assignment, where the connection probability is specified by an element of the $q \times q$ affinity matrix $\ket{\omega}$; the edge probability distribution of a vertex pair is given by the Bernoulli distribution. 
Thus, the likelihood of the stochastic block model is given as 
\begin{align}\label{SBMlikelihood}
p(A, \ket{\sigma} \lvert \ket{\omega}, \ket{\gamma}, q) 
&= p(A \lvert \ket{\sigma}, \ket{\omega}, q)p(\ket{\sigma}\lvert \ket{\gamma}, q) \notag\\
&= \prod_{i} \gamma_{\sigma_{i}} \prod_{i<j} \omega_{\sigma_{i}\sigma_{j}}^{A_{ij}} \left( 1 - \omega_{\sigma_{i}\sigma_{j}} \right)^{1-A_{ij}}. 
\end{align}
When a higher connection probability is provided for pairs of vertices with the same cluster assignment (as compared to pairs of vertices with different cluster assignments), i.e., $\omega_{\sigma\sigma} > \omega_{\sigma\sigma^{\prime}}$ ($\sigma \ne \sigma^{\prime}$), a set of random graphs with a community structure can be generated. 
Generating the stochastic block model is a forward problem, and community detection is its inverse problem, i.e., the inference of $\ket{\gamma}$, $\ket{\omega}$, and $\ket{\sigma}$. 

\subsection{Degree-corrected stochastic block model with restricted model-parameter space}
Whereas the standard stochastic block model is very flexible, it is often not suitable to fit real-world networks, mainly because it can only have a binomial degree distribution, which is not true in many datasets. 
To resolve this problem, the so-called degree-corrected stochastic block model \cite{KarrerNewman2011} was proposed. 
Following Ref.~\cite{KarrerNewman2011}, the Bernoulli distribution for the edge probability of each vertex pair was approximated with the Poisson distribution, which is justified when the network is sparse. 
In this model, it is assumed that the mean of the Poisson distribution depends on the degrees of the vertex pair as well as on the affinity matrix. 
Hence, for a given affinity matrix $\ket{\omega}$ and the number of clusters $q$, the likelihood of the degree-corrected stochastic block model is given as 
\begin{align}\label{DCSBMlikelihood}
p(A, \ket{\sigma} \lvert \ket{\omega}, q) 
&= p(A \lvert \ket{\sigma}, \ket{\omega}, q)p(\ket{\sigma}\lvert q) \notag\\
&= \prod_{i<j} (d_{i}\omega_{\sigma_{i}\sigma_{j}}d_{j})^{A_{ij}} \mathrm{e}^{-d_{i}\omega_{\sigma_{i}\sigma_{j}}d_{j}}, 
\end{align}
where $d_{i}$ is the degree of vertex $i$. 
Here and hereafter, the uniform prior distribution for $p(\ket{\sigma})$ was considered for simplicity. 
Moreover, we neglected the existence of self-loops, which is also justified when the network is sparse. 
The log likelihood  reads 
\begin{align}\label{DCSBMloglikelihood}
\log p(A, \ket{\sigma} \lvert \ket{\omega}) 
&= \sum_{i<j} \left[ A_{ij} \log \left( d_{i} \omega_{\sigma_{i}\sigma_{j}} d_{j} \right) - d_{i} \omega_{\sigma_{i}\sigma_{j}} d_{j} \right], 
\end{align}
where we neglected a constant term. 

While the stochastic block model of Eq.~(\ref{DCSBMloglikelihood}) is able to express various modular structures, hereafter, we restrict our interest to the community structure. 
The affinity matrix $\ket{\omega}$ is then restricted to the form 
\begin{align}
\omega_{\sigma\sigma^{\prime}} 
=\begin{cases}
& \omega_{\mathrm{in}} \hspace{20pt} ( \sigma = \sigma^{\prime} ), \\
& \omega_{\mathrm{out}} \hspace{20pt} ( \sigma \ne \sigma^{\prime} ). 
\end{cases} \label{ParameterRestriction}
\end{align}
In other words, only whether a vertex pair is assigned to the same cluster or not is distinguished.  
This restriction to the inference algorithm using BP was proposed in Ref.~\cite{ZhangMoore2014}. 
One of the reasons why this restriction is employed is because it can considerably reduce the computational cost, so that performance on large networks with many clusters can be evaluated. 
Another reason is because some assessment criteria compared in this paper are specialized to the community structure, and their generalizations to general modular structures are not known. 

The log likelihood Eq.~(\ref{DCSBMloglikelihood}) is then simplified to  
\begin{align}
\log p(A, \ket{\sigma} \lvert \ket{\omega}) 
&= \sum_{i<j} \delta_{\sigma_{i}\sigma_{j}} \biggl[ 
A_{ij} \log \frac{\omega_{\mathrm{in}}}{\omega_{\mathrm{out}}} 
- (\omega_{\mathrm{in}} - \omega_{\mathrm{out}}) d_{i}d_{j} \biggr] \notag\\
&\hspace{20pt}+ \sum_{i<j} \biggl[ A_{ij} \log \left( \omega_{\mathrm{out}}d_{i}d_{j} \right) - \omega_{\mathrm{out}}d_{i}d_{j} \biggr]. 
\end{align}
Note that the second sum does not depend on the cluster assignment $\ket{\sigma}$. 
We consider this stochastic block model for community detection.

\section{Community detection algorithms}\label{Algorithm}
\subsection{Statistical inference}\label{StatisticalInference}
The goal of community detection is to determine the set of cluster assignments $\ket{\sigma}$; it is a hidden variable, and when the model parameter $\ket{\omega}$ is learned for a given number of clusters $q$, fitting the network for the stochastic block model is carried out by maximizing the marginal log likelihood or equivalently, minimizing the free energy,  
\begin{align}
f(\ket{\omega}, q) &= -\log\sum_{\ket{\sigma}} p(A, \ket{\sigma} \lvert \ket{\omega}, q). \label{FreeEnergy}
\end{align}
Unfortunately, this optimization problem is computationally difficult in general, and a number of approximate methods have been proposed in the literature. 
The EM algorithm is employed in this paper, which is a popular method for fitting the stochastic block model. 
To obtain the minimum of the free energy, the EM algorithm iteratively optimizes the distribution of the hidden variable $\ket{\sigma}$ with a fixed model parameter $\ket{\omega}$ (E step) and the optimization of $\ket{\omega}$ with a fixed distribution of $\ket{\sigma}$ (M step). 
For the E step, we use the BP algorithm which will be explained later in this section. 
Thus, we obtain the probability distribution of the cluster assignment for each vertex, such that Eq.~(\ref{FreeEnergy}) is expected to be minimized.  
Hereafter, we often omit the number of clusters $q$ in the argument, which is always given as an input; we try various values of $q$ for model assessment.

When the affinity matrix $\ket{\omega}$ is fixed as a constant in the E step, the free energy reads  
\begin{align}\label{MarginalLikelihood}
f(\ket{\omega}, q) 
&= \text{const.} - \log \sum_{\ket{\sigma}} \mathrm{e}^{2L \beta Q(\ket{\sigma})}, 
\end{align}
where 
\begin{align}
& Q(\ket{\sigma}) = \frac{1}{2L} \sum_{i<j} \delta_{\sigma_{i}\sigma_{j}} \biggl[ 
A_{ij} - \alpha \frac{d_{i}d_{j}}{2L} \biggr], \label{Modularity}\\
& \alpha = \frac{2L}{\beta}(\omega_{\mathrm{in}} - \omega_{\mathrm{out}}), 
\hspace{10pt} 
\beta = \log \frac{\omega_{\mathrm{in}}}{\omega_{\mathrm{out}}}, 
\end{align}
are the modularity function $Q(\ket{\sigma})$, resolution parameter $\alpha$ \cite{Reichardt2006}, and inverse temperature $\beta$, respectively. 
This indicates that modularity maximization can be regarded as a special case of the inference using the stochastic block model; the partition with the maximum modularity coincides with the result of the statistical inference when the entropic effect is ignored, or $\beta\to\infty$. This is known as the maximum \textit{a posteriori} (MAP) estimate \cite{Nishimori2001}. 
The connection between likelihood maximization and modularity maximization was first discussed in Ref.~\cite{Newman2013} for $q=2$ in the context of spectral graph partitioning; the above relation was pointed out in Ref.~\cite{ZhangMoore2014}, which discusses a finite temperature formulation of the modularity maximization. 
It is known that $\beta$ also plays the role of a resolution parameter \cite{Shulke2015} that controls the typical scale of clusters.

In Refs.~\cite{ZhangMoore2014,Shulke2015}, $\alpha$ is set to unity and $\beta$ is treated as an input parameter, which corresponds to fitting a network with a fixed affinity matrix. 
However, it is more natural to learn them instead. 
The learning of $\omega_{\mathrm{in}}$ and $\omega_{\mathrm{out}}$ can be carried out in a straightforward manner. 
They are obtained as the values that minimize the free energy, Eq.~(\ref{MarginalLikelihood}). 
The derivatives with respect to the model parameters \cite{FootnoteEMalgorithm} yield 
\begin{align}
\hat{\omega}_{\mathrm{in}} &= 
\frac{\sum_{(i,j) \in E} \bracket{\delta_{\sigma_{i}\sigma_{j}}} }{\sum_{i<j} d_{i}d_{j} \bracket{\delta_{\sigma_{i}\sigma_{j}}} }, \label{hatomegain}\\
\hat{\omega}_{\mathrm{out}} &= 
\frac{\sum_{(i,j) \in E} \left( 1 - \bracket{\delta_{\sigma_{i}\sigma_{j}}}\right) }{\sum_{i<j} d_{i}d_{j} \left( 1 - \bracket{\delta_{\sigma_{i}\sigma_{j}}}\right) }, \label{hatomegaout}
\end{align}
where $\delta_{\sigma \sigma^{\prime}}$ is the Kronecker delta, $\bracket{\cdots}$ is the average with respect to the current estimate of the posterior distribution $p(\ket{\sigma} \lvert A, \ket{\omega})$ and the hat notation indicates the estimated quantity. 
Let $n_{\sigma} = \sum_{i} \bracket{\delta_{\sigma\sigma_{i}}}$ denote the number of nodes within cluster $\sigma$. 
As mentioned in Ref.~\cite{ZhangMartinNewman2015}, if we assume that $p(\ket{\sigma} \lvert A, \ket{\omega})$ is the distribution that prevents $n_{\sigma}$ from fluctuating significantly, i.e., $\bracket{n_{\sigma}^{2}} \approx \bracket{n_{\sigma}}^{2}$, 
\begin{align}
\sum_{i<j}d_{i}d_{j} \bracket{\delta_{\sigma_{i}\sigma_{j}}}
&\approx \frac{1}{2} \sum_{\sigma} \bracket{\sum_{i} d_{i}\delta_{\sigma\sigma_{i}}}^{2}. 
\end{align}
We also assumed that the overcounting for $i=j$ in the sum is negligible. 
Then, Eqs.~(\ref{hatomegain}) and (\ref{hatomegaout}) can be approximated as 
\begin{align}
\hat{\omega}_{\mathrm{in}} &= 
2 \frac{\sum_{\sigma}\sum_{(i,j) \in E} \bracket{\delta_{\sigma\sigma_{i}}\delta_{\sigma\sigma_{j}}} }{\sum_{\sigma} \left(\sum_{i} d_{i} \bracket{\delta_{\sigma\sigma_{j}}}\right)^{2} }, \label{approxhatomegain}\\
\hat{\omega}_{\mathrm{out}} &= 
2 \frac{ L - \sum_{\sigma}\sum_{(i,j) \in E} \bracket{\delta_{\sigma\sigma_{i}}\delta_{\sigma\sigma_{j}}} }{ (2L)^{2} - \sum_{\sigma} \left(\sum_{i} d_{i} \bracket{\delta_{\sigma\sigma_{j}}}\right)^{2} }. \label{approxhatomegaout}
\end{align}
Note that the update of model parameters only costs linear time; therefore, it is not a bottleneck in the algorithm. 

To evaluate the probability of cluster assignment for each vertex, BP is used (see Refs.~\cite{MezardMontanari2009,Decelle2011,Decelle2011a,ZhangMoore2014} for details). 
The marginal probability $\psi^{i}_{\sigma_{i}}$ of vertex $i$'s cluster assignment $\sigma_{i}$ can be obtained by marginalizing the likelihood Eq.~(\ref{DCSBMlikelihood}). Using the tree approximation, which is valid for sparse networks, it can be expressed as 
\begin{align}
&\psi^{i}_{\sigma_{i}} 
\!=\! \frac{1}{Z^{i}} 
\!\!\prod_{k \in \partial i} \left[ \sum_{\sigma_{k}} \psi^{k \to i}_{\sigma_{k}} \mathrm{e}^{\beta \delta_{\sigma_{i}\sigma_{k}}} \right] 
\!\!\prod_{\ell \notin i \cup \partial i} \left[ \sum_{\sigma_{\ell}} \psi^{\ell \to i}_{\sigma_{\ell}} \mathrm{e}^{-\alpha\beta \frac{d_{i} d_{\ell}}{2L} \delta_{\sigma_{i}\sigma_{\ell}} } \right] \notag\\
&\!\!=\! \frac{1}{Z^{i}} 
\!\!\prod_{k \in \partial i} \left[ 1 + \psi^{k \to i}_{\sigma_{i}} \left( \mathrm{e}^{\beta} - 1 \right) \right] 
\!\!\prod_{\ell \notin i \cup \partial i} \!\! \left[ 1 + \psi^{\ell \to i}_{\sigma_{i}} \left( \mathrm{e}^{-\alpha\beta \frac{d_{i} d_{\ell}}{2L}} - 1\right) \right], \label{PSIi}
\end{align}
where $\partial i$ denotes the neighboring vertices of $i$.  
In Eq.~(\ref{PSIi}), $\psi^{k \to i}_{\sigma_{i}}$ indicates the cavity bias from vertex $k$ to vertex $i$, that is, the marginal probability of $k$ without the marginalization from $i$, and $Z^{i}$ is the normalization factor. 
Assuming that $\alpha\beta d_{i} d_{\ell}/2L = O(N^{-1})$, we can further approximate $\psi^{i}_{\sigma_{i}}$ and $\psi^{i \to j}_{\sigma_{i}}$ as 
\begin{align}
\psi^{i}_{\sigma_{i}} 
&\approx \frac{1}{Z^{i}} \mathrm{e}^{-\frac{\alpha\beta}{2L} d_{i} \theta_{\sigma_{i}}}
\prod_{k \in \partial i} \left[ 1 + \psi^{k \to i}_{\sigma_{i}} \left( \mathrm{e}^{\beta} - 1 \right) \right], \label{PSIi-cav}\\
\psi^{i \to j}_{\sigma_{i}} 
&\approx \frac{1}{Z^{i \to j}} \mathrm{e}^{-\frac{\alpha\beta}{2L} d_{i} \theta_{\sigma_{i}}}
\prod_{k \in \partial i \backslash j} \left[ 1 + \psi^{k \to i}_{\sigma_{i}} \left( \mathrm{e}^{\beta} - 1 \right) \right], \label{PSIi-full}
\end{align}
respectively, where $\theta_{\sigma} \equiv \sum_{\ell} d_{\ell} \psi^{\ell}_{\sigma}$. 
The cavity bias $\psi^{i \to j}_{\sigma_{i}}$ is normalized by $Z^{i \to j}$ (see Ref.~\cite{ZhangMoore2014} for details). 
Using these quantities, we can estimate the elements in Eqs.~(\ref{approxhatomegain}) and (\ref{approxhatomegaout}) as 
\begin{align}
&\sum_{i} d_{i} \bracket{\delta_{\sigma\sigma_{i}}} = \sum_{i} d_{i} \psi^{i}_{\sigma} = \theta_{\sigma}, \\
&\sum_{\sigma} \bracket{\delta_{\sigma\sigma_{i}}\delta_{\sigma\sigma_{j}}} 
= \sum_{\sigma} \frac{1}{Z^{ij}} \psi^{i \to j}_{\sigma} d_{i}\omega_{\mathrm{in}}d_{j} \psi^{j \to i}_{\sigma} \notag\\
&= \frac{\omega_{\mathrm{in}} \sum_{\sigma}\psi^{i \to j}_{\sigma}\psi^{j \to i}_{\sigma}}
{ \left(\omega_{\mathrm{in}} - \omega_{\mathrm{out}}\right) \sum_{\sigma} \psi^{i \to j}_{\sigma} \psi^{j \to i}_{\sigma} + \omega_{\mathrm{out}} } \hspace{10pt} \text{for } (i,j) \in E. \label{DiagSum2ptProb}
\end{align}
It should be noted that the iteration of Eqs.~(\ref{approxhatomegain}), (\ref{approxhatomegaout}), (\ref{PSIi-cav}), and (\ref{PSIi-full}) does not minimize the free energy Eq.~(\ref{FreeEnergy}) itself, but minimizes its approximated quantity called the Bethe free energy. We show the specific form of the Bethe free energy in Sec.~\ref{Assessment:BetheFreeEnergy}. 

The critical values of $\beta$ for the stochastic block model have been discussed in Refs.~\cite{ZhangMoore2014,Shulke2015}. 
There are three phases of state, depending on the value of $\beta$ and the strength of the community structure: the retrieval phase, paramagnetic phase, and spin-glass phase. 
In the retrieval phase, the fixed point of BP with the minimum Bethe free energy correctly indicates the community structure. 
In the paramagnetic phase, BP converges to the so-called factorized state as the minimum of the Bethe free energy. 
In the factorized state, for any vertex $i$, the marginal probability distribution of the cluster assignment $\psi^{i}_{\sigma}$ has a uniform distribution, $\psi^{i}_{\sigma} = 1/q$. In other words, any vertex has an equal probability of joining any cluster.  Therefore, the resulting partition does not exhibit any community structure. 
Finally, the spin-glass phase is the phase in which BP typically does not converge. This is also the case in which the statistically significant community structure cannot be retrieved. 
In the case of the standard stochastic block model with equal size clusters, for a given number of clusters $q^{\ast}$, the critical value of $\beta$ between the paramagnetic phase and the spin-glass phase obtained by the stability of the factorized state against a random perturbation is 
\begin{align}
\beta^{\ast} &= \log \left( \frac{q^{\ast}}{\sqrt{c}-1} + 1 \right), \label{betaast}
\end{align}
where $c$ is the average degree. 
The lower bound estimate of $\beta$ that prevents BP from going into the paramagnetic phase is given by: 
\begin{align}
\beta_{0} = \log \left( \frac{q^{\ast}}{c-1} + 1 \right). \label{betazero}
\end{align}
In practice, it cannot be uniquely determined whether BP belongs to the retrieval phase, paramagnetic phase, or spin-glass phase, because real-world networks do not precisely emulate the stochastic block model. 
However, they work as the reference values of $\beta$ to obtain an intuition regarding which phase BP belongs to. 
In Ref.~\cite{ZhangMoore2014}, it is suggested that $\beta = \beta^{\ast}$ should be used as an input, because BP is expected to belong to the retrieval phase with this value.

The effect due to the absence of model-parameter learning can be interpreted as follows. 
Given that the model only distinguishes whether a pair of vertices is in the same cluster or not, the specific values of $\omega_{\mathrm{in}}$ and $\omega_{\mathrm{out}}$ may not be so crucial for the resulting cluster assignment. 
Conversely, when other statistical quantities such as likelihood or cross-validation errors are considered, erroneous model-parameter estimates may cause a large bias. 
As we observe in Sec.~\ref{ComparativeAnalysis}, the results of the criteria that depend only on cluster assignments (e.g., modularity and minimum description length of the map equation) are not very sensitive to model-parameter learning, while the criteria that utilize the model parameters (e.g., the Bethe free energy and cross-validation errors) are ill-behaved without learning.

\subsection{Greedy algorithms}
In the previous section, the free energy minimization based on the stochastic block model has been considered. 
In the limit of $\beta \to \infty$, it reduces to the maximization of the modularity function $Q(\ket{\sigma})$ in Eq.~(\ref{Modularity}), or the energy minimization. 
In this case, the probability distribution with respect to $\ket{\sigma}$ is no longer considered, and our goal here is to find the best cluster assignment for each vertex. 

While a number of algorithms have been proposed in the literature, perhaps, greedy algorithms, such as the Louvain method \cite{Blondel2008}, are the most widely used in practice. 
Another greedy algorithm for community detection that we analyze is the Infomap \cite{Rosvall2008}, which optimizes the map equation \cite{Rosvall2008} (see Sec.~\ref{Assessment:MapEquation} for the details of the map equation). 
In such algorithms, we assign a unique cluster label for each vertex at the beginning, i.e., $q=N$, and merge and update their assignments as referring to the neighboring vertices to achieve a higher or lower value of the objective function, e.g., $Q(\ket{\sigma})$. 
Note that the number of clusters is also determined automatically during the optimization process. 
Although these greedy algorithms are extremely fast, as we will observe in Sec.~\ref{Comparison:Greedy}, they tend to largely overfit when the algorithm is trapped in a local extremum of the energy landscape. 
The situation is very severe particularly when the landscape is glassy \cite{Good2010}.

\subsection{Spectral methods}\label{Algorithm:SpectralMethods}

\begin{figure}[t]
 \begin{center}
   \includegraphics[width=0.99 \columnwidth, bb=15 9 373 171]{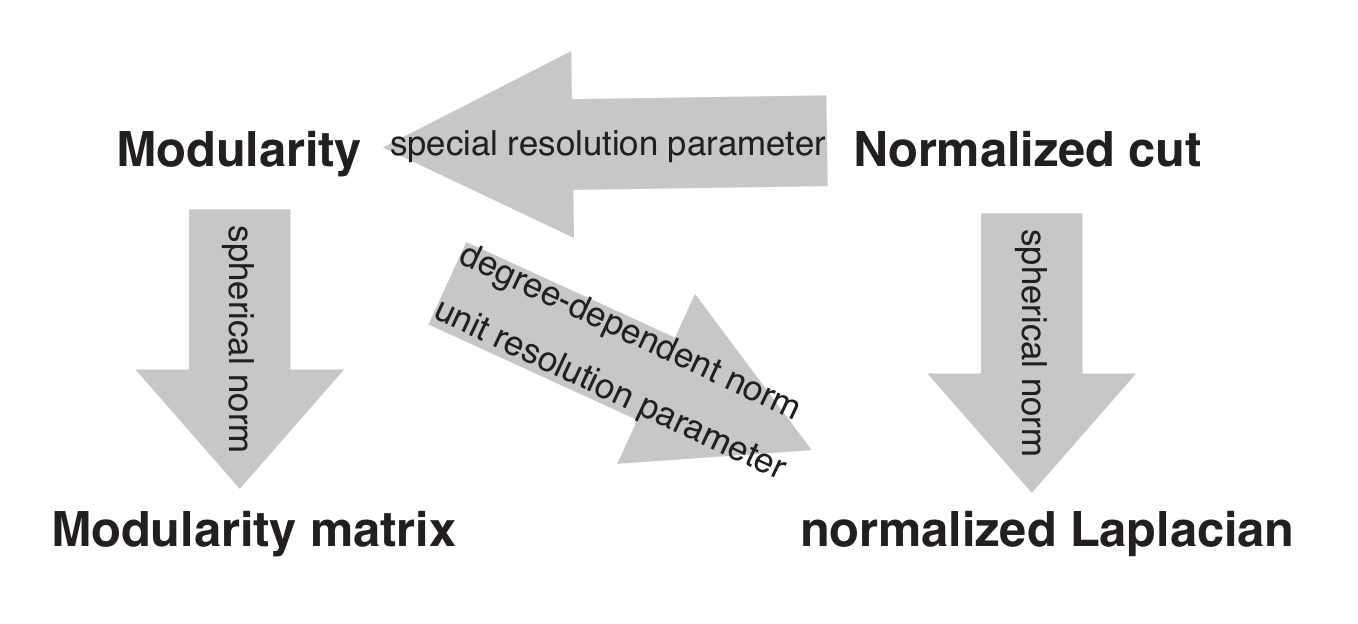}
 \end{center}
 \caption{Relationships with the modularity function, normalized cut, and the corresponding spectral methods in the case of bisection. }
 \label{SpectralRelationship}
\end{figure}

Other commonly used algorithms are spectral methods. 
In this section, the focus is on the case of $q=2$ (bisection). 
Let us consider the case of modularity maximization. 
While maximizing $Q(\ket{\sigma})$ is originally a discrete optimization problem, the assignments $\ket{\sigma}$ are relaxed to a real vector $\ket{x} \in \mathbb{R}^{N}$ with a spherical normalization constraint, i.e., 
\begin{align}
\max_{\ket{x}} \, Q(\ket{x}) \hspace{20pt} \sum_{i} x^{2}_{i} = N. \label{SpectralModularity-1}
\end{align}
Here, $Q(\ket{x}) = \bra{x} \mathcal{Q} \ket{x}$ and $\mathcal{Q}$ is a matrix the element of which is given as 
\begin{align}
\mathcal{Q}_{ij} = A_{ij} - \alpha\frac{d_{i}d_{j}}{2L}. \label{SpectralModularity-2}
\end{align}
This matrix is known as the modularity matrix \cite{Newman2006politicalbooks,Newman2006PRE}. 
If Eq.~(\ref{SpectralModularity-1}) is rewritten using the Lagrange multiplier, an eigenvalue problem is obtained with respect to $\mathcal{Q}$ and the leading eigenvector is expected to be correlated to the optimum assignments. 
Although $\mathcal{Q}$ is a dense matrix, because it is due to the rank 1 matrix of the second term in Eq.~(\ref{SpectralModularity-2}), its leading eigenvalues and eigenvectors can be computed efficiently using the power iteration \cite{Github-Modularity}.

A more classical version of a spectral method is the one based on the minimization of an objective function called the normalized cut \cite{Shi2000}. 
The normalized cut $f_{\mathrm{Ncut}}$ is defined as, for $\sigma_{i} \in \{1,2\}$, 
\begin{align}
f_{\mathrm{Ncut}}(\ket{\sigma}) &\propto \frac{\sum_{i,j} A_{ij} \delta_{\sigma_{i},1} \delta_{\sigma_{j},2}}{(\sum_{i (\sigma_{i} = 1)} d_{i})(\sum_{j (\sigma_{j} = 2)} d_{j})}. 
\end{align}
Analogous to the case of modularity, the continuous relaxation with the spherical normalization constraint yields the eigenvalue problem with respect to the normalized Laplacian $\mathcal{L}$, which is defined as: 
\begin{align}
\mathcal{L} &= I - D^{-1/2}AD^{-1/2}, 
\end{align}
where $I$ is the identity matrix and $D^{k} \equiv \mathrm{diag}\{ d^{k}_{1}, \dots, d^{k}_{N}\}$ (see Ref.~\cite{Luxburg2007} for details). 

The optimizations of the normalized cut and modularity look different. However, it is known that minimizing the normalized cut $f_{\mathrm{Ncut}}(\ket{\sigma})$ is equivalent to maximizing modularity $Q(\ket{\sigma})$ with a special choice of the resolution parameter $\alpha$ \cite{KawamotoKabashimaEPL2015} at the level of discrete optimization. Moreover, when the problem is relaxed to a continuous optimization, it is also possible to formulate the modularity maximization as the eigenvalue problem of the normalized Laplacian \cite{Newman2013}; it can be done by imposing a degree-dependent normalization constraint instead of the spherical normalization constraint and setting the resolution parameter $\alpha$ to unity. These relationships are summarized in Fig.~\ref{SpectralRelationship}. 

In general, the leading $q$ eigenvectors are expected to be correlated to the optimum $q$-way partition. Thus, to determine the assignment of each vertex, those eigenvectors have to be rounded, e.g., using the K-means method. However, the focus of this paper is only on the eigenvalues, because they are sufficient to estimate the number of clusters. 

While the above two spectral methods are based on energy minimization, a spectral method related to the Bethe free energy minimization was proposed in Ref.~\cite{Krzakala2013}. 
The matrix that appears in this method is called the non-backtracking matrix $\mathcal{B}$, and is derived from the linear stability analysis of the BP algorithm that minimizes the Bethe free energy. 
The non-backtracking matrix $\mathcal{B}$ is not a symmetric matrix, and its specific form is given as: 
\begin{align}
\mathcal{B} = 
\left(\begin{array}{cc}
0 & D- I \\
-I & A\\
\end{array}\right). \label{NBmatrix}
\end{align}
A symmetric variant of the non-backtracking matrix is also proposed in Ref.~\cite{SaadeBetheHessian}. 

In Sec.~\ref{Assessment:SpectralMethods}, we will discuss the properties of the spectra of the above matrices and  their use as the assessment criteria of the number of clusters.

\section{Assessment criteria of the number of clusters}\label{ModelAssessmentCriteria}
In this section, we explain the assessment criteria of the number of clusters $q^{\ast}$. 
To determine it using an algorithm in which $q$ is given as an input, we assess the quality of the clustering based on a criterion, as we sweep the value of $q$. 
It should be noted that, although the input value of the number of clusters, namely, the maximum number of clusters that the vertices can be assigned to, is $q$, the resulting partition might have less than $q$ clusters.

\subsection{Bethe free energy}\label{Assessment:BetheFreeEnergy}
In the present framework of statistical inference, the most natural assessment is to measure the free energy and observe its saturation as an increment of the number of clusters $q$. 
When the network is generated by a block model with $q^{\ast}$ clusters, the marginal likelihood will not be increased for $q>q^{\ast}$. 
Therefore, we expect that a parsimonious number of clusters can be selected from the saturation of the free energy. 

As mentioned above, the algorithm using BP does not minimize the free energy itself. 
Instead, it minimizes the Bethe free energy as an approximated quantity, and it can be written in terms of the cavity bias $\psi^{i \to j}_{\sigma}$ and affinity matrix $\ket{\omega}$ as follows. 
\begin{align}
f_{\mathrm{Bethe}} &= - \frac{1}{\beta N} \left( \sum_{i} \log Z^{i} - \sum_{(i,j)\in E} \log Z^{ij} - \sum_{(i,j)\notin E} \log \tilde{Z}^{ij} \right), 
\end{align}
where 
\begin{align}
Z^{i} &= \sum_{\sigma} \mathrm{e}^{-d_{i} h_{\sigma_{i}}} 
\prod_{k \in\partial i} \Biggl[ \sum_{\sigma_{k}} \psi_{\sigma_{k}}^{k \to i} 
d_{k}\omega_{\sigma_{k}\sigma_{i}}d_{i}
 \Biggr], \\
Z^{ij} &= \sum_{\sigma\sigma^{\prime}} 
\psi^{i \to j}_{\sigma} 
d_{i} \omega_{\sigma\sigma^{\prime}} d_{j} 
\psi^{j \to i}_{\sigma^{\prime}} & \text{for } (i,j) \in E, \\
\tilde{Z}^{ij} &= \sum_{\sigma\sigma^{\prime}} 
\psi^{i \to j}_{\sigma} 
(1 - d_{i} \omega_{\sigma\sigma^{\prime}} d_{j} )
\psi^{j \to i}_{\sigma^{\prime}} & \text{for } (i,j) \notin E, \\
h_{\sigma} &= \sum_{k} d_{k} \sum_{\sigma_{k}} \psi^{k}_{\sigma_{k}} \omega_{\sigma_{k}\sigma}. 
\end{align}
Some simple algebra shows that the Bethe free energy here $f_{\mathrm{Bethe}}$ is related to the Bethe free energy in Ref.~\cite{ZhangMoore2014} (which we refer to as $f^{\mathrm{mod}}_{\mathrm{Bethe}}$) as 
\begin{align}
f_{\mathrm{Bethe}} &= f^{\mathrm{mod}}_{\mathrm{Bethe}} + C(\omega_{\mathrm{out}}), \\
C(\omega_{\mathrm{out}}) &= -\frac{c}{2\beta}\left( 2L \omega_{\mathrm{out}} + \log\omega_{\mathrm{out}} + \frac{1}{L}\sum_{i} d_{i} \log d_{i} \right). 
\end{align}

\subsection{Modularity}\label{Assessment:Modularity}
While modularity appeared as an objective function with a fixed $q$ in Sec.~\ref{StatisticalInference}, it was originally defined as an assessment criterion of the number of clusters \cite{Newman2004}. 
In modularity, the strength of a community structure is measured by comparing the actual network and a randomized network in each cluster. 
Although the performance of modularity is not considered state of the art, it has been extensively studied and used as a baseline in many benchmark tests. 

Precisely speaking, while the sum is taken over every vertex pair $(i, j)$ ($i<j$) in Eq.~(\ref{Modularity}), the sum is taken over all possible combinations of vertices (including the case $i=j$) in the original definition, although this does not cause a qualitative difference unless the self-loops are significant. 
The modularity function of Eq.~(\ref{Modularity}) with the partition obtained by free energy minimization is sometimes distinguished as the retrieval modularity. However, it is referred to as \textit{modularity} in this paper for simplicity. 
The partition is selected with a maximum modularity, or the parsimonious one among the partitions with a high modularity. 

From the view point of statistical inference, the modularity maximization corresponds to a maximum likelihood estimate; i.e., there is no penalty term. 
In principle, it can still assess the number of clusters because the degrees of freedom of the affinity matrix $\ket{\omega}$ are restricted as in Eq.~(\ref{ParameterRestriction}), and thus, the model with a larger $q$ does not contain the model with a smaller $q$ as a subset. 
In addition, if we tune the resolution parameter $\alpha$, the likelihood varies, and the optimum value $q^{\ast}$ changes.

\subsection{The map equation}\label{Assessment:MapEquation}
Another popular criterion is the map equation \cite{Rosvall2008,Rosvall2011}, in which the strength of the community structure is measured in terms of the minimum description length of a random walker. 
The map equation encodes the moves of a random walker on a given network using multiple codebooks. 
Specifically, it considers a codebook that encodes moves between clusters, as well as codebooks that encode moves within each cluster. 
Given that the codewords of different codebooks can be overlapped, a proper assignment of clusters will compress the description length of a random walker. Moreover, by using the codebooks of superclusters, i.e., the clusters of clusters, its hierarchical extension can be performed naturally. 
The map equation also has an interesting feature in that it allows for the consideration of flow information, e.g., the directedness of edges, although we do not address this point in this paper (see Refs.~\cite{Rosvall2008,Rosvall2011} for more details). 

The excellent performance of the map equation and its greedy implementation (Infomap) has been shown in numerous articles. 
As with modularity, one can use the minimum description length of the map equation for model assessment only and perform community detection based on another objective function. 
It should be noted that the characterization of a cluster in the map equation is not equivalent to that of the stochastic block model. 
However, when densely connected components exist in a network, the minimum description length of a random walker is further compressed by clustering them; thus, it is expected that an optimal partition in the sense of modularity is also a good partition in the sense of the map equation. 

It is also debatable whether we should consider the hierarchical nature of the map equation \cite{Rosvall2011}. 
The map equation is naturally formulated as a hierarchical clustering, and the fundamental two-level method can be regarded as a truncation of the general multilevel method. 
Nevertheless, we measure the minimum description length of the two-level method and compare it with other model assessment criteria, because it is not always possible to measure the minimum description length in the sense of the multilevel map equation. 
Whereas the multilevel map equation assumes a hierarchical structure, for example, in the case of partitions using the inference algorithm considered in this paper, each partition with different values of $q$ is independent and is not constrained to constitute a hierarchical structure.

\subsection{Spectral band}\label{Assessment:SpectralMethods}
\begin{figure}[t]
 \begin{center}
   \includegraphics[width=1 \columnwidth, bb=0 0 499 296]{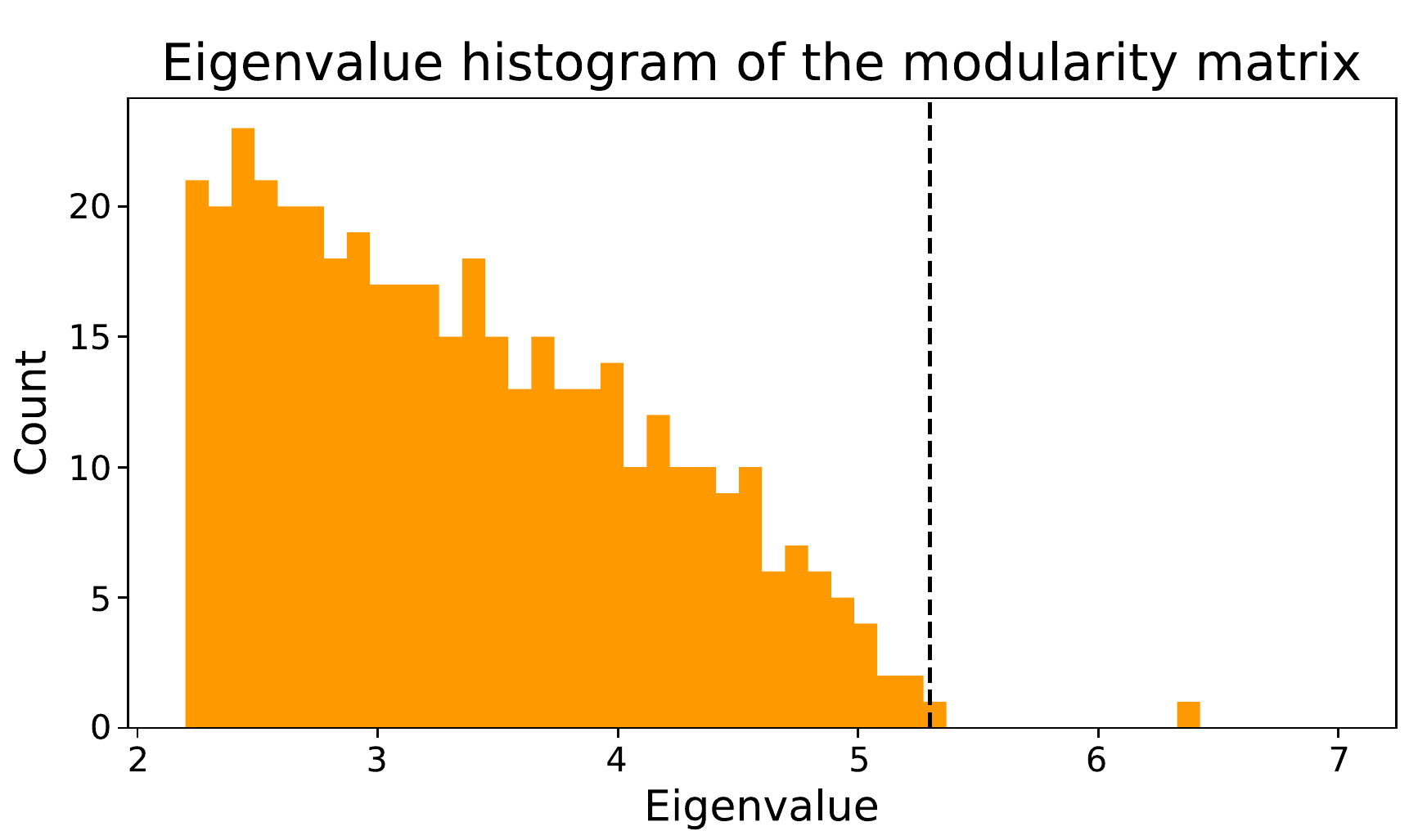}
 \end{center}
 \caption{(Color online) A part of the histogram of the eigenvalues of the modularity matrix for the standard stochastic block model. 
 The network has $N=2000$ with two equally sized planted clusters and an average degree equal to 6 . The dashed line indicates the estimate of the boundary of the spectral band calculated by the result in Ref.~\cite{KawamotoKabashimaEPL2015}.}
 \label{SpectralDensity}
\end{figure}

The spectra of matrices $\mathcal{Q}$, $\mathcal{L}$, and $\mathcal{B}$, in Sec.~\ref{Algorithm:SpectralMethods} can be used to estimate the number of clusters. 
In the case of a uniform random graph, in the infinite graph size limit, the spectrum of a corresponding matrix exhibits a non-zero spectral density within a finite range. 
In other words, the spectral band can be observed, as exemplified in Fig.~\ref{SpectralDensity}; it is often referred to as the \textit{semicircle law} \cite{MehtaRMT} in the case of a symmetric matrix.  
The spectral band stems purely from the random nature of a network, and if a characteristic structure in a network exists, the eigenvalues outside of the spectral band, i.e., isolated eigenvalues, will be observed. 
As we mentioned in Sec.~\ref{Algorithm:SpectralMethods}, because the leading eigenvectors are expected to be correlated to the optimum partition, the number of statistically significant clusters can be estimated by counting the isolated eigenvalues. 

While it is sometimes possible to evaluate the boundary of the spectral band by visual inspection, it is not trivial and it is preferable to have its estimate. 
The estimate of the boundary of the spectral band of the modularity matrix $\mathcal{Q}$ was first derived in Ref.~\cite{NadakuditiNewman2012}, the estimate was then generalized for random networks with arbitrary expected degrees \cite{NadakuditiNewman2013}, and arbitrary degree sequences \cite{ZhangNadakuditiNewman2014}. 
These results, however, assume that the average degree is sufficiently large. 
A result that is applicable to sparse networks is derived in Ref.~\cite{KawamotoKabashimaEPL2015}; although it is still a mean-field result, it yields the estimate for a random network with an arbitrary degree sequence and it is exact when the network is regular. 

Although the boundary estimate of the spectral band of the normalized Laplacian $\mathcal{L}$ is also possible using the mean-field approximation \cite{KawamotoKabashimaPRE2015}, it is known that $\mathcal{L}$ seriously suffers from the emergence of localized eigenvectors; those localized eigenvectors consist of a few elements with very large values and most of the elements are close to zero. 
These localized eigenvectors are not correlated to the optimum partition and can deteriorate the estimate of the number of clusters. 
For this reason, we do not pursue the assessment using the spectrum of $\mathcal{L}$ in this paper. 

On the other hand, the non-backtracking matrix $\mathcal{B}$ tends to avoid the emergence of the localized eigenvectors, and its spectral band has a clear boundary at $\sqrt{\rho(\mathcal{B})}$ \cite{Saade2014}, where $\rho(\mathcal{B})$ is the spectral radius of $\mathcal{B}$. 
As seen in Sec.~\ref{Comparison:Inference}, the assessment using the non-backtracking matrix performs well in many cases. 
Note that, however, the non-backtracking matrix is not completely free from the localized eigenvectors \cite{Kawamoto_JSTAT2016,Pastor-Satorras2016}. 


\subsection{Prediction errors}\label{Assessment:PredictionErrors}
Finally, we explain the cross-validation estimates of prediction errors, which are also useful to estimate the number of clusters (see Ref.~\cite{KawamotoKabashimasbmBIX} for the detailed derivations). 
Although evaluating cross-validation errors is computationally demanding in general, the leave-one-out cross-validation (LOOCV) is an exceptional case and the corresponding errors can be obtained efficiently using the result of BP \cite{KawamotoKabashimasbmBIX}. 

We consider four types of cross-validation errors, the Bayes prediction error, Gibbs prediction error, MAP estimate of the Gibbs prediction error, and Gibbs training error. 
We refer to $A^{\cav{i}{j}}$ as the adjacency matrix without the knowledge of $A_{ij}$. 
Given $A^{\cav{i}{j}}$, the cluster assignment probability of $i$ and $j$ is 
\begin{align}
p(\sigma_{i}, \sigma_{j} \lvert A^{\cav{i}{j}}) = \psi^{i \to j}_{\sigma} \psi^{j \to i}_{\sigma^{\prime}}. 
\end{align}
Then, the prediction probability $\hat{p}(A_{ij} = 1 | A^{\cav{i}{j}})$ that $i$ and $j$ are connected is: 
\begin{align}
& \hat{p}(A_{ij} = 1 \lvert A^{\cav{i}{j}}) \notag\\
&= \sum_{\sigma_{i}, \sigma_{j}} \hat{p}(A_{ij} = 1 \lvert \sigma_{i}, \sigma_{j}) p(\sigma_{i}, \sigma_{j} \lvert A^{\cav{i}{j}}) \notag\\
&= \sum_{\sigma\sigma^{\prime}} 
\psi^{i \to j}_{\sigma} 
d_{i} \omega_{\sigma\sigma^{\prime}} d_{j} 
\psi^{j \to i}_{\sigma^{\prime}}
= Z^{ij}. 
\end{align}
Note that the two-point partition function $Z^{ij}$ is the normalization factor in Eq.~(\ref{DiagSum2ptProb}) and is not equivalent to the two-point partition function defined in Ref.~\cite{ZhangMoore2014}, which does not have a probabilistic interpretation. 
The cross-entropy error function with respect to $\hat{p}(A_{ij} \lvert A^{\cav{i}{j}})$, which is referred to as the Bayes prediction error of LOOCV, $E_{\mathrm{Bayes}}$, is: 
\begin{align}
E_{\mathrm{Bayes}}(q) 
&= -\frac{1}{L} \sum_{i<j} \biggl[ A_{ij} \log \hat{p}(A_{ij} = 1 \lvert A^{\cav{i}{j}}) \notag\\
&+ (1-A_{ij}) \log \left(1 - \hat{p}(A_{ij} = 1 \lvert A^{\cav{i}{j}})\right) \biggr]. 
\end{align}
Using the fact that $\omega_{\sigma\sigma^{\prime}} = O(N^{-1})$, it can be approximated as: 
\begin{align}
E_{\mathrm{Bayes}}(q) 
&\simeq 1-\frac{1}{L} \sum_{(i,j) \in E} \log Z^{ij}, 
\end{align}
where we neglected the $O(N^{-1})$ term. 
The Bayes prediction error $E_{\mathrm{Bayes}}$ should be the appropriate choice for assessing models in terms of the predictability of edges when the network is generated by the stochastic block model. 
However, this is often not the case.  
Hence, the Gibbs prediction error $E_{\mathrm{Gibbs}}$ is considered, which is a rough estimate of the prediction error compared to $E_{\mathrm{Bayes}}$. 
While the probability with respect to $\sigma_{i}$ and $\sigma_{j}$ is marginalized when the cross-entropy error function is measured in $E_{\mathrm{Bayes}}$, a specific choice is made regarding $\sigma_{i}$ and $\sigma_{j}$ first, and the average is taken later in $E_{\mathrm{Gibbs}}$. 
Thus, we have: 
\begin{align}\label{EGibbs}
& E_{\mathrm{Gibbs}}(q) \notag\\
&\simeq 1 - \frac{1}{L} \sum_{(i,j)\in E} \sum_{\sigma_{i}, \sigma_{j}} p(\sigma_{i}, \sigma_{j} \lvert A^{\cav{i}{j}}) \log\left( \hat{p}(A_{ij} = 1 \lvert \sigma_{i}, \sigma_{j}) \right) \notag\\
&= 1-\frac{1}{L} \sum_{(i,j)\in E} \sum_{\sigma_{i}, \sigma_{j}} \psi^{i \to j}_{\sigma_{i}}  \psi^{j \to i}_{\sigma_{j}} \log \left(d_{i}\omega_{\sigma_{i} \sigma_{j}}d_{j}\right) \notag\\
&= -\frac{\beta}{L} \sum_{(i,j)\in E} \sum_{\sigma} \psi^{i \to j}_{\sigma}\psi^{j \to i}_{\sigma} - \log\omega_{\mathrm{out}} + \mathrm{const.}, 
\end{align}
where we again neglected the $O(N^{-1})$ term. 
By replacing $\psi^{i \to j}_{\sigma}$ with the delta function that has a peak at $\mathrm{argmax}_{\sigma} \psi^{i \to j}_{\sigma}$, the MAP estimate of the Gibbs prediction error is obtained, which is referred to as $E_{\mathrm{MAP}}$. 

The Gibbs training error $E_{\mathrm{training}}$ can be derived in the same manner. 
In $E_{\mathrm{training}}$, we include the information of $A_{ij}$ for the probability with respect to $\sigma_{i}$ and $\sigma_{j}$. 
Thus, we have, 
\begin{align}
&E_{\mathrm{training}}(q) \notag\\
&\simeq 1 - \frac{1}{L} \sum_{(i,j)\in E} \sum_{\sigma_{i}, \sigma_{j}} p(\sigma_{i}, \sigma_{j} \lvert A) \log\left( \hat{p}(A_{ij} = 1 \lvert \sigma_{i}, \sigma_{j}) \right) \notag\\
&= 1-\frac{1}{L} \sum_{(i,j)\in E} \sum_{\sigma_{i}, \sigma_{j}} \frac{\psi^{i \to j}_{\sigma_{i}} d_{i}\omega_{\sigma_{i} \sigma_{j}}d_{j} \psi^{j \to i}_{\sigma_{j}}}{Z^{ij}} \log \left(d_{i}\omega_{\sigma_{i} \sigma_{j}}d_{j}\right) \notag\\
&= \mathrm{const.} -\frac{1}{L} \sum_{(i,j)\in E} \notag\\
&\frac{\omega_{\mathrm{out}}\log\omega_{\mathrm{out}} + \beta \left(\omega_{\mathrm{out}} + (\alpha/2L) \log\omega_{\mathrm{out}}\right) \sum_{\sigma} \psi^{i \to j}_{\sigma}\psi^{j \to i}_{\sigma}}
{\omega_{\mathrm{out}} + (\alpha\beta/2L) \sum_{\sigma} \psi^{i \to j}_{\sigma}\psi^{j \to i}_{\sigma}}. 
\end{align}
Again, the $O(N^{-1})$ term was neglected.

Note that the complexity of computing the Bethe free energy and the cross-validation errors is considerably reduced by restricting the parameter space of the stochastic block model. 
While the stochastic block model required a computation of $O(q^{2}L)$ in the general case, it is $O(L)$ with the restriction: Eq.~(\ref{ParameterRestriction}).

\section{Comparative analysis}\label{ComparativeAnalysis}
In this section, a comparative analysis of the assessment of the number of clusters was conducted using synthetic and real-world networks. 
For the synthetic networks, the planted number of clusters is denoted as: $q_{\mathrm{planted}}$. 

\subsection{Assessment using the greedy algorithms}\label{Comparison:Greedy}

\begin{figure}[t!]
 \begin{center}
   \includegraphics[width=0.9 \columnwidth, bb=18 19 558 965]{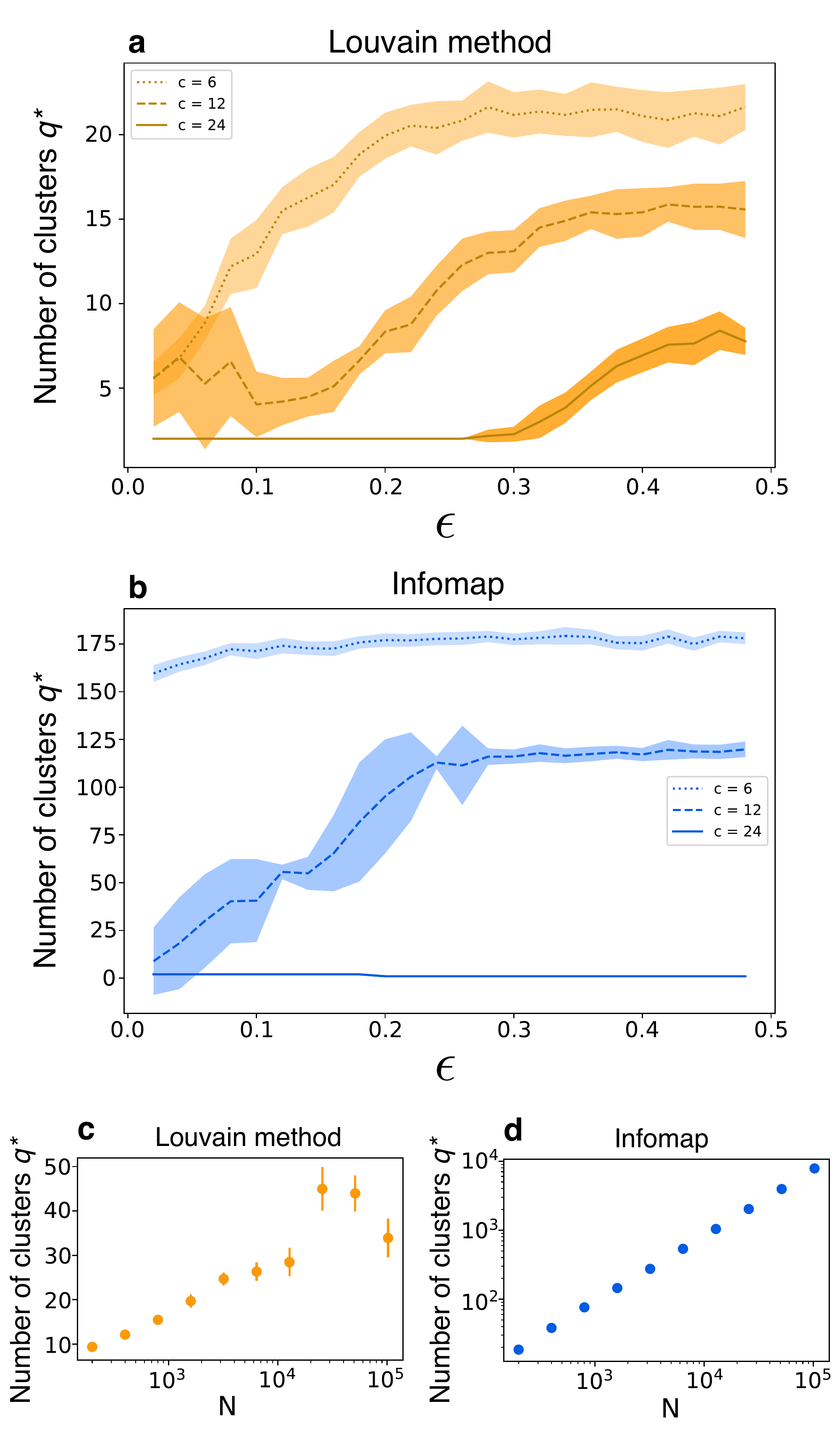}
 \end{center}
 \caption{(Color online) Numbers of clusters $q^{\ast}$ detected by the (a) Louvain method and (b) Infomap for instances of the standard stochastic block model. Each network has two equally sized clusters as a planted structure, and the networks are generated for various values of $\epsilon = \omega_{\mathrm{out}}/\omega_{\mathrm{in}}$, i.e., the strength of community structure. 
The algorithms were executed 30 times for each network; the resulting number of clusters fluctuates depending on the initial cluster assignments, and the shaded regions show the standard deviations from the mean value. In each case, the stochastic block models of $N=2,000$ with the average degrees $c=6$, $12$, and $24$ are evaluated.
The bottom figures show the $N$ dependence on the estimated number of clusters $q^{\ast}$ for the (c) Louvain method and (d) Infomap; the planted structure is of two equally sized clusters with $c = 6$ and $\epsilon = 0.5$, and the experimental procedure is the same as in Figs.~\ref{LouvainInfomapAssessment}{\bf a} and \ref{LouvainInfomapAssessment}{\bf b}. The error bars indicate the standard deviations.}
 \label{LouvainInfomapAssessment}
\end{figure}

\begin{figure}[t!]
 \begin{center}
   \includegraphics[width=0.9 \columnwidth, bb=2 0 400 602]{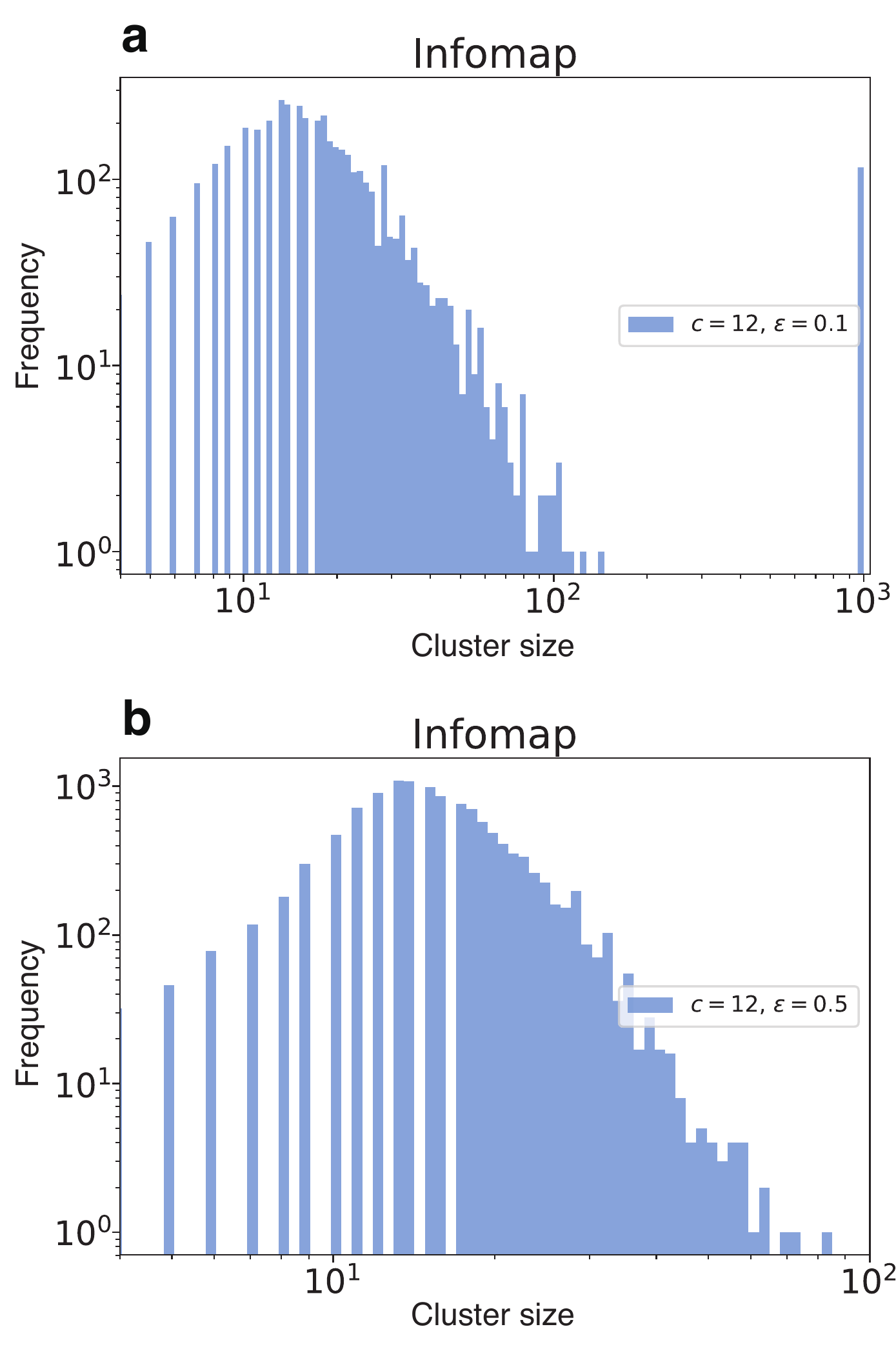}
 \end{center}
 \caption{(Color online) Histograms of the cluster size distributions detected by the Infomap for the same stochastic block models as in Fig.~\ref{LouvainInfomapAssessment} with (a) $c = 12$ and $\epsilon = 0.1$ and (b) $c = 12$ and $\epsilon = 0.5$. 
 The algorithm is run $100$ times on the same network and the results show their cumulative frequencies. 
 When the community structure is strong ($\epsilon = 0.1$), one of the planted clusters seems to be detected. 
 Conversely, only small clusters that are broadly distributed are detected when $\epsilon = 0.5$.
Note that the specific form of the cluster size distribution possibly depends on the details of the algorithm. 
 }
 \label{InfomapClusterSizeDistribution}
\end{figure}

The performance of the greedy algorithms was first examined on the basis of the standard stochastic block model. 
Figures \ref{LouvainInfomapAssessment}{\bf a} and \ref{LouvainInfomapAssessment}{\bf b} show the number of clusters detected using the Louvain method and the two-level Infomap, respectively. 
The horizontal axes represent the strength of the community structure $\omega_{\mathrm{out}}/\omega_{\mathrm{in}} \equiv \epsilon$. 
The Louvain method is a hierarchical clustering algorithm that aims to optimize modularity, while the (two-level) Infomap is a non-hierarchical clustering algorithm that aims to optimize the map equation. 
For the implementation, we used Ref.~\cite{louvain-igraph} for the Louvain method and Ref.~\cite{infomap-igraph} for the Infomap. 

All instances considered in this section have $q_{\mathrm{planted}} = 2$. 
Given that the stochastic block model is exactly the model assumed in the inference algorithm, the assessment by the Bethe free energy and some of the prediction errors are known to be very accurate \cite{Mossel2014,Massoulie2014,KawamotoKabashimasbmBIX}, even when the planted modular structure is very weak. 

When the average degree is sufficiently high and the community structure is strong (i.e., $\epsilon \sim 0$), both algorithms correctly detect two clusters. 
However, when the networks are sparse and the community structure is weak (i.e., $\epsilon \gg 0$), those algorithms tend to largely overfit. 
Moreover, as shown in Figs.~\ref{LouvainInfomapAssessment}{\bf c} and \ref{LouvainInfomapAssessment}{\bf d}, the detected number of clusters increases as the network becomes larger. 
A non-hierarchical clustering algorithm for modularity \cite{Clauset_FastGreedy,fastgreedy-igraph} and the multilevel Infomap \cite{Rosvall2011} were also tested. 
Although the tendency that the hierarchical clusterings slightly prevent overfitting was confirmed, significant differences were not observed. 

It should be noted that, detecting too many clusters does not readily imply the failure of the algorithm. 
For example, when the result consists of a few large clusters and many very small clusters, significant clusters can be extracted via a visual inspection. 
This is actually the case for instances with strong community structures. 
Otherwise, the sizes of clusters can be broadly distributed, and such a visual inspection may fail. 
Such situations are exemplified in Fig.~\ref{InfomapClusterSizeDistribution} for the Infomap. 

\begin{figure*}[t!]
 \begin{center}
   \includegraphics[width=1.99 \columnwidth, bb=4 5 724 346]{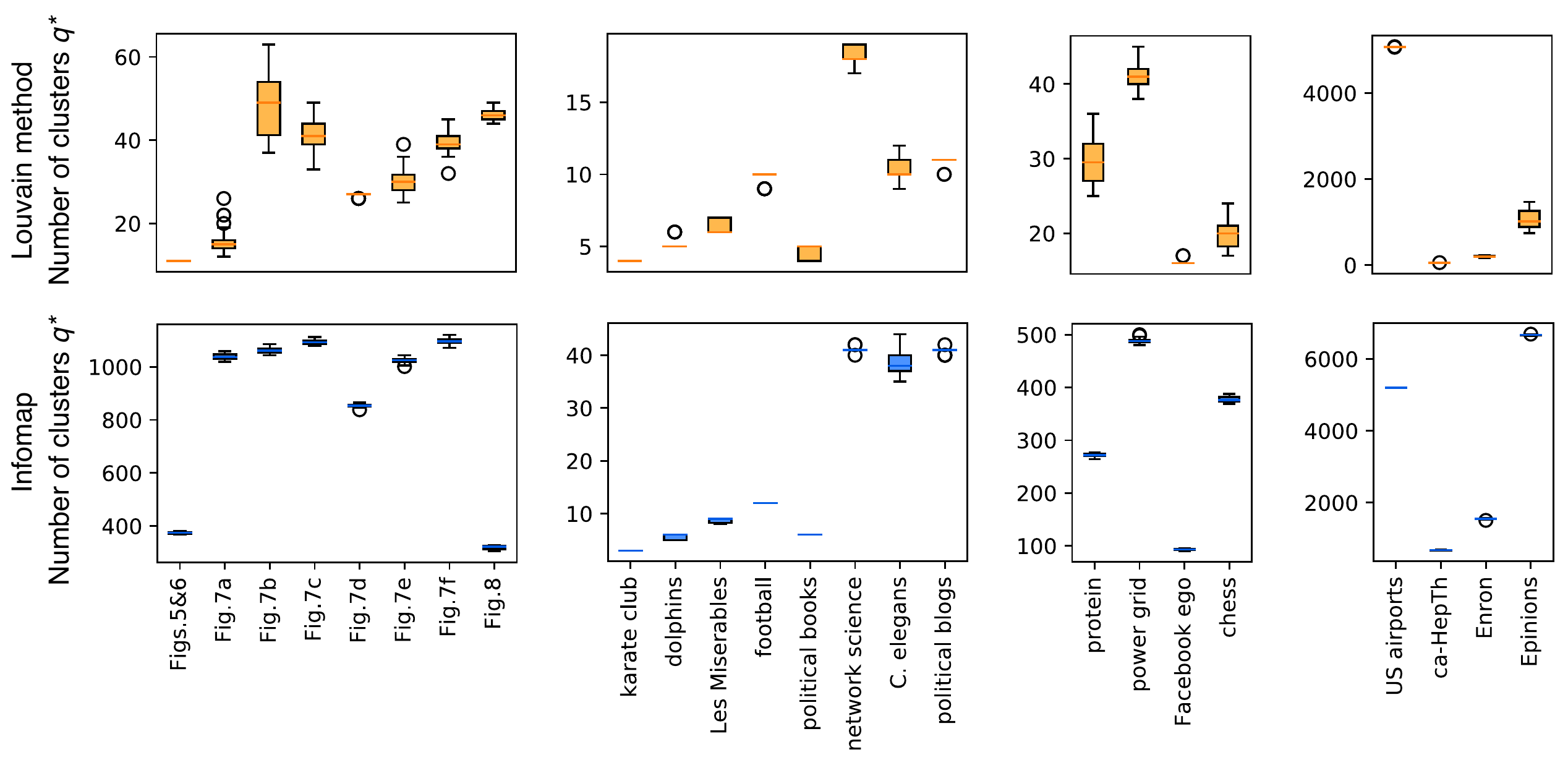}
 \end{center}
 \caption{(Color online) Box plots of the estimates of the number of clusters $q^{\ast}$ using the Louvain method (top) and the two-level Infomap (bottom) on the synthetic and real-world networks that are analyzed in Secs.~\ref{Comparison:Inference} and \ref{Comparison:Spectral}. 
 The algorithms were executed 30 times for each network. 
 Each box represents the range from the upper quantile to the lower quantile, the line in the box represents the median, whiskers from the box represent the upper and lower extremes, and circles represent the outliers, which are significantly far from the upper and lower quantiles. 
}
 \label{LouvainInfomapKstatistics}
\end{figure*}

As a reference to the comparative analysis of the latter sections, we list the results of the greedy algorithms on synthetic and real-world networks in Fig.~\ref{LouvainInfomapKstatistics}. 
The descriptions of the networks can be found in Sec.~\ref{Comparison:Inference} and the references therein. 

\subsection{Assessment using the inference algorithm}\label{Comparison:Inference}

\begin{table}[t]
     \begin{center}
     \begin{tabular}{lcccccccccc}
     \toprule
      Figure & $c$ & $d_{\mathrm{max}}$ & $\mu$ & $\tau_{1}$ & $\tau_{2}$ & $N^{\sigma}_{\mathrm{min}}$ & $N^{\sigma}_{\mathrm{max}}$ & $q_{\mathrm{planted}}$ & $q^{\ast}_{\mathrm{mod}}$ & $q^{\ast}_{\mathrm{NBT}}$\\ 
      \midrule
      \ref{LFR1Learned} \& \ref{LFR1NotLearned} & $9.9$ & $960$ & $0.1$ & $2$ & $1$ & $991$ & $1064$ & $11$ & $8$ & $15$ \\
      \ref{LFR5}{\bf a} & $6.5$ & $100$ & $0.3$ & $2$ & $1$ & $633$ & $946$ & $13$ & $12$ & $12$ \\
      \ref{LFR5}{\bf b} & $6.5$ & $100$ & $0.5$ & $2$ & $1$ & $633$ & $946$ & $13$ & $8$ & $13$ \\
      \ref{LFR5}{\bf c} & $6.5$ & $100$ & $0.8$ & $2$ & $1$ & $633$ & $946$ & $13$ & $1$ & $1$ \\
      \ref{LFR5}{\bf d} & $6.5$ & $100$ & $0.3$ & $2$ & $1$ & $110$ & $796$ & $27$ & $21$ & $27$ \\
      \ref{LFR5}{\bf e} & $6.5$ & $100$ & $0.5$ & $2$ & $1$ & $110$ & $796$ & $27$ & $12$ & $25$ \\
      \ref{LFR5}{\bf f} & $6.6$ & $98$ & $0.8$ & $2$ & $1$ & $119$ & $965$ & $23$ & $1$ & $1$ \\
      \ref{LFR6} & $7.4$ & $991$ & $0.3$ & $1.8$ & $1$ & $122$ & $887$ & $26$ & $2$ & $27$
      \\ \bottomrule
      \end{tabular}
      \caption{Parameters of the LFR networks. All networks have $N \simeq 10^{4}$. The average degree $c$, maximum degree $d_{\mathrm{max}}$, size of the smallest cluster $N^{\sigma}_{\mathrm{min}}$, and size of the largest cluster $N^{\sigma}_{\mathrm{min}}$ are the realized values, while the mixing parameter $\mu$, minus the exponent of the degree distribution $\tau_{1}$, and minus the exponent of the planted cluster size distribution $\tau_{2}$ are the input values. 
      The planted cluster size distribution is controlled by: $N^{\sigma}_{\mathrm{min}}$ and $N^{\sigma}_{\mathrm{max}}$, while $\tau_{2}$ is fixed. 
      A comparative analysis for different values of $\tau_{2}$ would be difficult because a required graph size $N$ becomes extremely large for a slight change of $\tau_{2}$. 
      The planted number of clusters $q_{\mathrm{planted}}$ and the estimates using the modularity matrix $q^{\ast}_{\mathrm{mod}}$ and the non-backtracking matrix $q^{\ast}_{\mathrm{NBT}}$ are indicated in the last three columns.}
      \label{LFRtable}
      \end{center}
\end{table}

\begin{figure}[t!]
 \begin{center}
   \includegraphics[width=0.9 \columnwidth, bb=0 0 386 555]{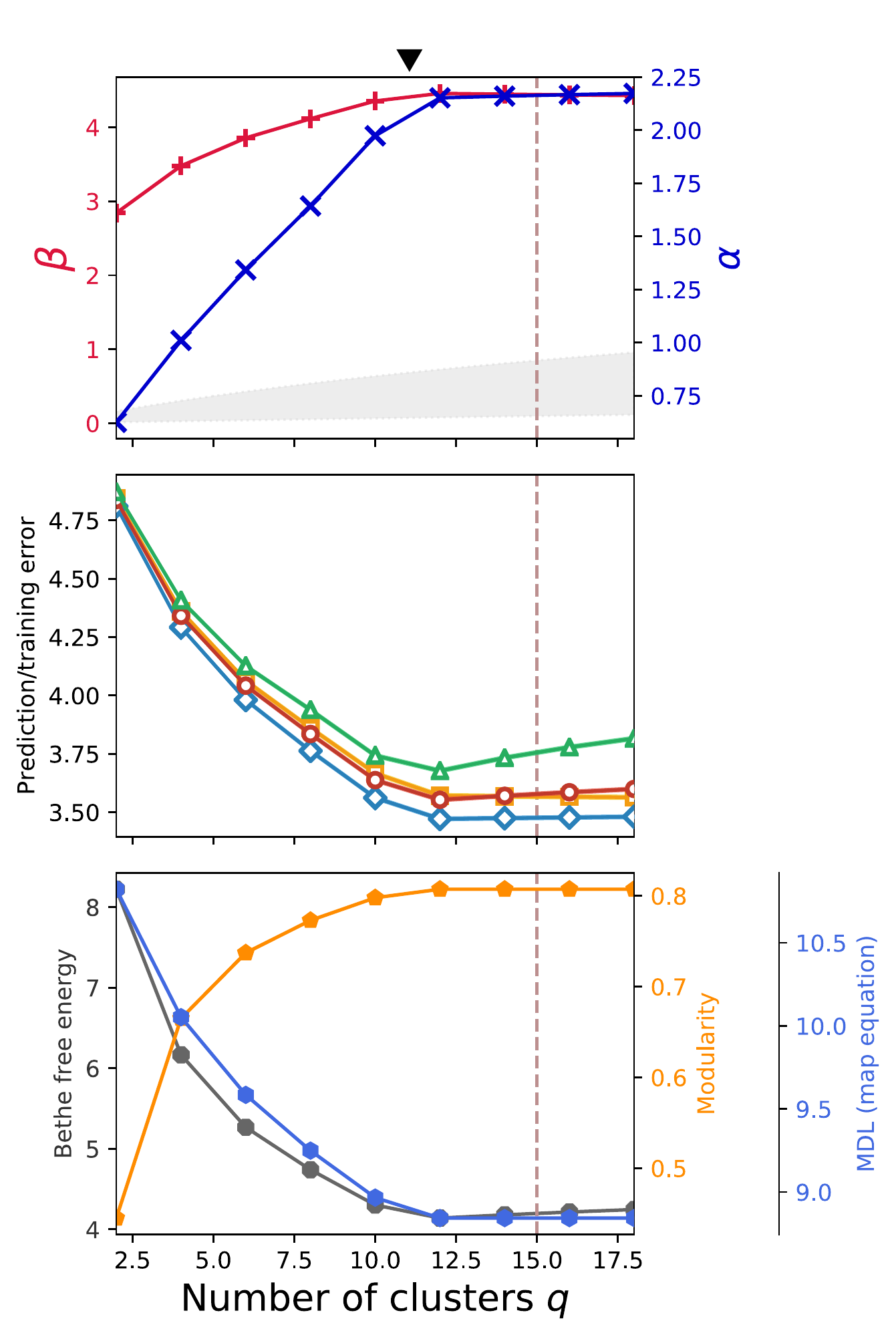}
 \end{center}
 \caption{(Color online) Assessment of various criteria with respect to given inputs of $q$ for a LFR network. 
	The top panel indicates the model parameters $\alpha$ (blue cross) and $\beta$ (red plus), where the shaded region indicates the region of $\beta$ between Eqs.~(\ref{betaast}) and (\ref{betazero}). 
	The middle panel indicates the Bayes prediction errors $E_{\mathrm{Bayes}}$ (red circles), 
	Gibbs prediction errors $E_{\mathrm{Gibbs}}$ (green triangles), 
	Gibbs training errors $E_{\mathrm{training}}$ (blue diamonds), and 
	MAP estimates $E_{\mathrm{MAP}}$ of $E_{\mathrm{Gibbs}}$ (yellow squares).
	The bottom panel indicates the modularity (yellow pentagon), map equation (blue hexagon), and Bethe free energy (gray octagon). 
	In each panel, the number of clusters selected by the spectral method of the non-backtracking matrix is indicated by a vertical dashed line. 
	The planted number of clusters $q_{\mathrm{planted}}$ is indicated with a filled triangle at the top of the figure. 
	}
 \label{LFR1Learned}
\end{figure}
\begin{figure}[t]
 \begin{center}
   \includegraphics[width=0.9 \columnwidth, bb=0 0 386 555]{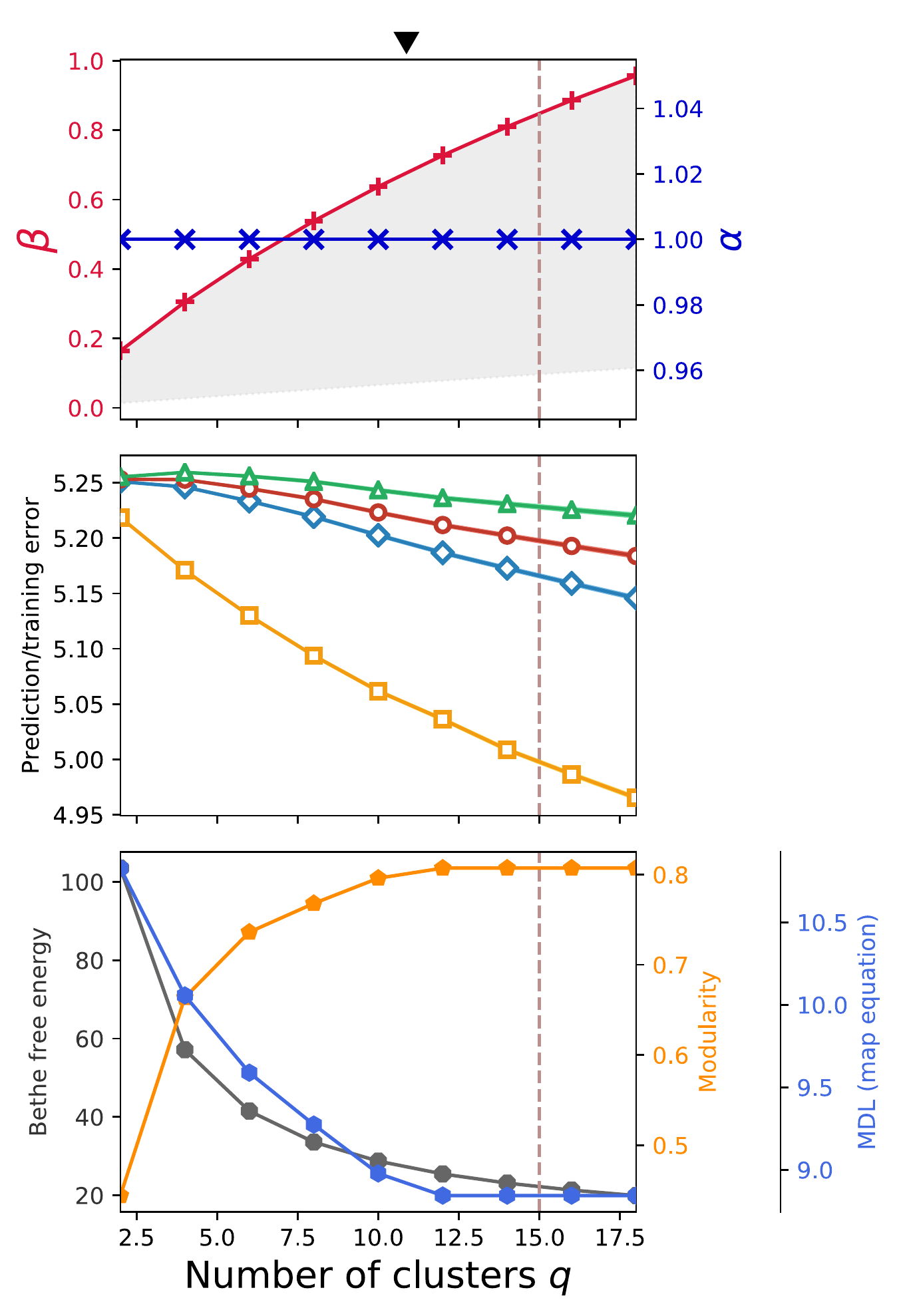}
 \end{center}
 \caption{(Color online) Result of the same LFR network as Fig.~\ref{LFR1Learned}. Here, statistical inference is performed without the model-parameter learning; the model parameters are fixed to: $\alpha=1$ and $\beta=\beta^{\ast}$. 
}
 \label{LFR1NotLearned}
\end{figure}

In this section, the performance of various assessment criteria based on the statistical inference algorithm that were explained in Sec.~\ref{StatisticalInference} is analyzed. 
First, the performance on synthetic networks, called the LFR network \cite{LFRbenchmark2009}, is analyzed. 
The LFR network is a random graph model that has a power-law degree distribution and, as a planted structure, a power-law cluster size distribution. 
The parameters of the LFR networks considered are listed in Table~\ref{LFRtable}. 
Although the LFR network is often analyzed as a random graph that emulates typical real-world networks, in this paper it is not argued whether the parameter set investigated is ``realistic'' or not. 
In fact, it is not obvious whether the LFR network really emulates typical real-world networks, because, as can be seen from Fig.~\ref{InfomapClusterSizeDistribution}, a broad cluster size distribution can be obtained fictitiously. 

Figure \ref{LFR1Learned} shows the result for an instance with a strong community structure (small mixing parameter $\mu$, in terms of the LFR network). 
Although the network has vertices with very large degrees, the cluster sizes are set to be almost equal. 
For this network, all the criteria support values close to $q_{\mathrm{planted}}$, although the criterion based on the non-backtracking matrix slightly overfits. 

While the values of the model parameters are learned in Fig.~\ref{LFR1Learned}, we show the result for the same network without the model-parameter learning in Fig.~\ref{LFR1NotLearned}. 
As we can see from Figs.~\ref{LFR1Learned} and \ref{LFR1NotLearned}, modularity and the map equation behave similarly in both cases. 
As we mentioned at the end of Sec.~\ref{StatisticalInference}, this may be due to the fact that the cluster assignment is not very sensitive to specific values of model parameters, at least when the network has a strong community structure. 
Conversely, the performance of the Bethe free energy and cross-validation errors change qualitatively, indicating that the learning step cannot be skipped. 
Note that skipping the learning step does not necessarily mean that it is computationally more efficient. With an incorrect choice of the affinity matrix $\ket{\omega}$, it will be more difficult to fit the network. 
It turns out that BP requires more sweeps until convergence. 
Therefore, it is more beneficial to optimize the model parameters. 
The rest of the results in this paper are generated according to the algorithm with model-parameter learning.

\begin{figure*}[t]
 \begin{center}
   \includegraphics[width=2 \columnwidth, bb=0 0 1157 1244]{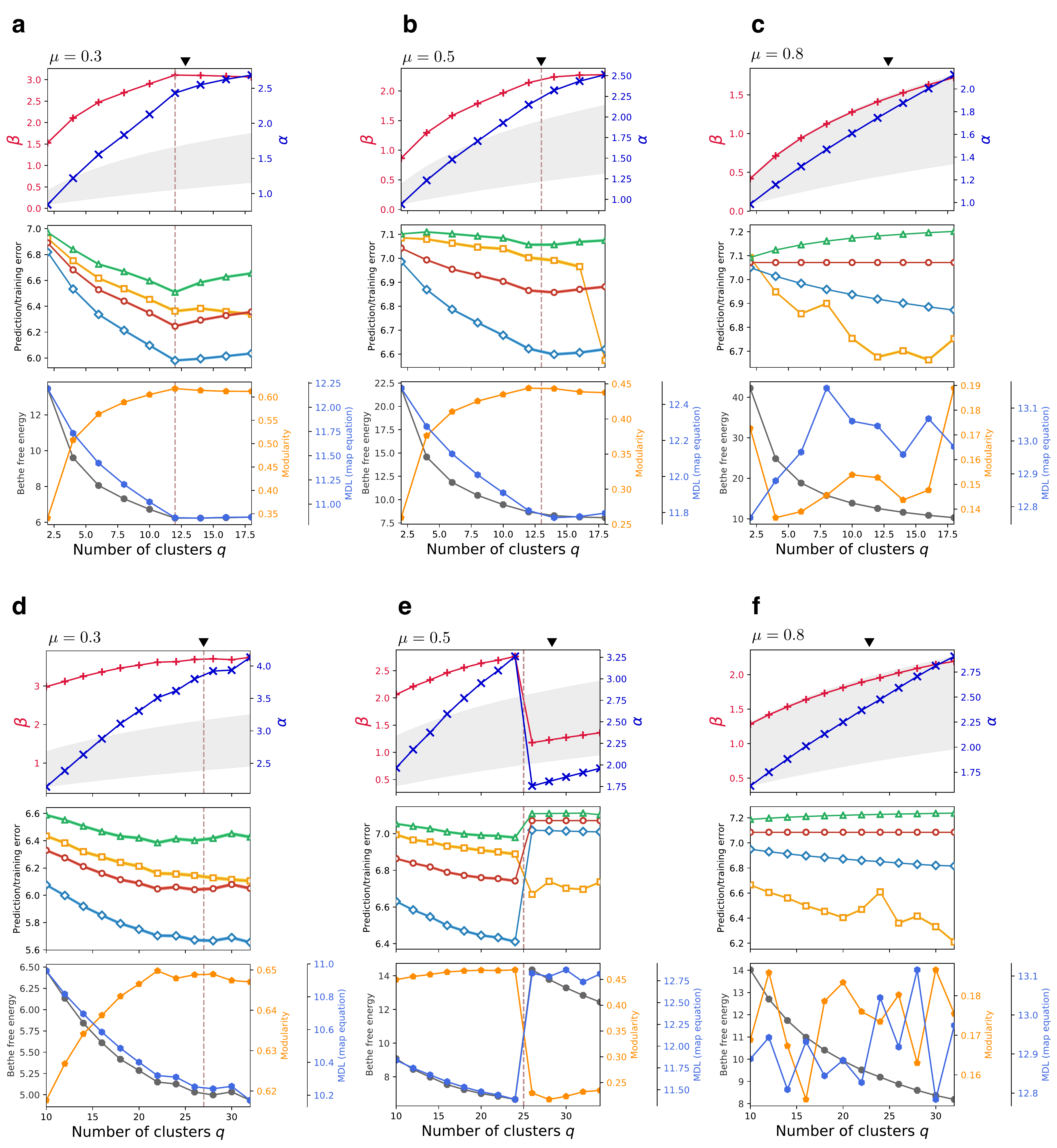}
 \end{center}
 \caption{(Color online) Assessment of various criteria with respect to given inputs of $q$ for several LFR networks. 
 They are plotted in the same manner as in Figs.~\ref{LFR1Learned} and \ref{LFR1NotLearned}. 
}
 \label{LFR5}
\end{figure*}

The LFR networks with weak community structures are shown in Fig.~\ref{LFR5}. 
Figures \ref{LFR5}{\bf a}--{\bf c} represent the results for networks with narrowly peaked planted cluster size distributions. 
Conversely, Fig.~\ref{LFR5}{\bf d}--{\bf f} represent the result for networks with broad planted cluster size distributions. 
Although it is difficult to thoroughly examine the effect of cluster size distribution, we can at least confirm that the performance of the present algorithm and assessment criteria are not very sensitive to the cluster size distribution. 

In the case of sparse networks such that the average degree is of $O(1)$, if the planted community structure is too weak, it becomes fundamentally impossible to retrieve the planted structure. In other words, the network becomes statistically impossible to distinguish from a uniform random graph. 
The critical strength of the community structure is called the detectability threshold, or the detectability limit \cite{Decelle2011a,MooreReview2017,FootnoteDetectabilityThreshold}. 
In terms of the spectral method, it is the case that the leading eigenvalues are buried in the spectral band. 
In terms of other assessment criteria, the slope of a validation curve becomes flat, or the value at $q=1$ becomes the minimum. 
In the case of the stochastic block model with equally sized clusters, this threshold is given by the value of $\epsilon$ that corresponds to $\beta^{\ast}$ in Eq.~(\ref{betaast}), and the paramagnetic phase will be observed beyond the detectability threshold. 

For the network in Fig.~\ref{LFR5}{\bf a}, all the criteria we consider behave reasonably, supporting the values close to $q_{\mathrm{planted}}$. 
For the network in Fig.~\ref{LFR5}{\bf b}, other than the Gibbs prediction error and its MAP estimate, the assessment criteria still support the values near $q_{\mathrm{planted}}$. 
Indeed, in the case of the stochastic block model, it is known that the Gibbs prediction error tends to underfit near the detectability threshold \cite{KawamotoKabashimasbmBIX}. 
Although the value of the (information-theoretic) detectability threshold for the LFR network is not known, the network in Fig.~\ref{LFR5}{\bf c} may be beyond the detectability threshold. The Gibbs prediction error and Bayes prediction error are minimized or saturated already at $q=1$ (not shown in the figure). The Bethe free energy exhibits a monotonic behavior, while the values of other criteria behave violently; this implies that the landscapes of the objective functions are glassy. 

More importantly, while we observed in Figs.~\ref{LouvainInfomapAssessment} and \ref{LouvainInfomapKstatistics} that the estimates by modularity and the map equation largely overfit when the greedy algorithm is used, the results in Fig.~\ref{LFR5} indicate that those criteria behave reasonably when statistical inference is used. 
Therefore, the shortcomings that we observed in Fig.~\ref{LouvainInfomapAssessment} were not the flaw of the criteria themselves, but of the greedy algorithms and the MAP estimate (i.e., $\beta\to\infty$) framework. 
In fact, when the assumed model is correct, it is known that the MAP estimate overfits compared to the estimate with the optimum inverse temperature $\beta$ \cite{ZhangMoore2014,MooreReview2017,Peixoto2017tutorial}. 
The contribution of this paper is that it confirms that the overfitting occurs with the greedy algorithms near the detectability threshold. 


While the results in Figs.~\ref{LFR5}{\bf d} and \ref{LFR5}{\bf f} are similar to those in Figs.~\ref{LFR5}{\bf a}--{\bf c}, the result in Fig.~\ref{LFR5}{\bf e} is qualitatively different. 
As the input values of $q$ are increased, at some point, BP converged to the factorized state \cite{FootnoteFactorizedState}; as a result, the estimated value of $\beta$ jumps, and some prediction errors become constant. 
Convergence to the factorized state is a desirable feature of BP; it implies that BP has reached the detectability threshold and that there is no significant structure. 
Note, however, that it is often difficult to determine whether BP is actually in the paramagnetic phase or the retrieval phase. 
Given that the factorized state always exists as a fixed point of BP in the retrieval phase, it is possible that BP is trapped in a local minimum of the Bethe free energy, while the correct initial state would converge to the global minimum. 

\begin{figure}[t!]
 \begin{center}
   \includegraphics[width=0.9 \columnwidth, bb=0 0 386 555]{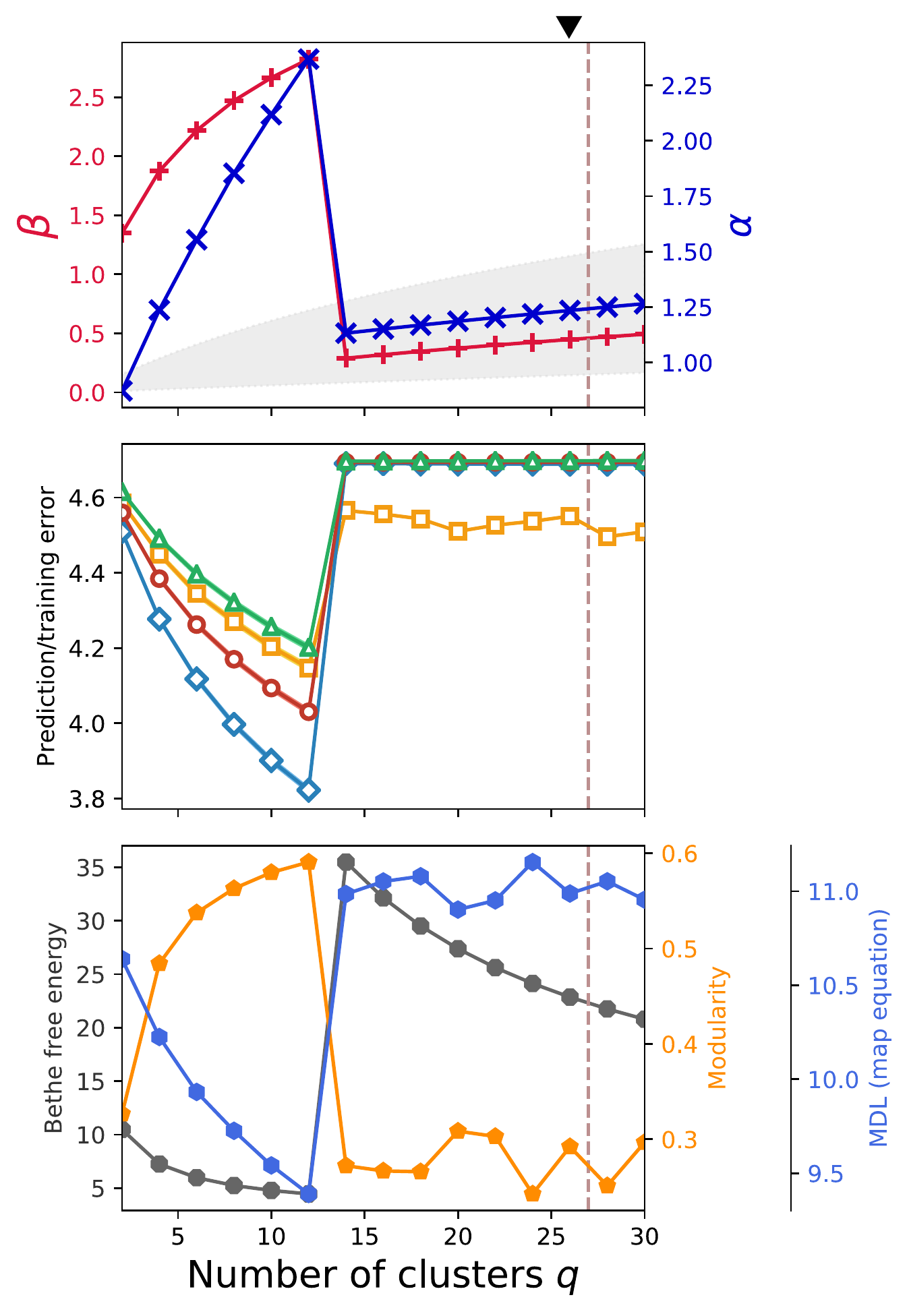}
 \end{center}
 \caption{(Color online) Assessment of various criteria with respect to given inputs of $q$ for a LFR network. 
 They are plotted in the same manner as in Figs.~\ref{LFR1Learned}--\ref{LFR5}. 
}
 \label{LFR6}
\end{figure}

As the final example using the LFR network, consider the case in which the assessment seems to fail because of the EM algorithm. 
Consider a network that has a broad degree distribution as in Figs.~\ref{LFR1Learned} and \ref{LFR1NotLearned}, in addition to a weak community structure and a broad planted cluster size distribution. 
As shown in Fig.~\ref{LFR6}, while the spectrum of the non-backtracking matrix exhibits the estimate of $q^{\ast}$ near $q_{\mathrm{planted}}$, such estimates cannot be obtained via the other criteria, because BP converges to the factorized state at $q=12$, although the values of the criteria significantly vary when they reach this value. 
In this case, we can hardly conclude that there are no statistically significant structures beyond $q=12$, and it is more natural to conclude that the BP converged to a local minimum of the Bethe free energy. 
Note that, even if we accept the estimate of $q^{\ast} = 27$, we cannot obtain a result with 27 clusters; recall that the input value of $q$ is the maximum number of clusters allowed, and the actual number of clusters that can be obtained is much less than 27. 
Readers might wonder what factor dominates the performance of the EM algorithm in the LFR network. 
Although the degree distribution seems to be an important factor, because there are many model parameters in the LFR network, it is difficult to precisely determine parameter dependencies experimentally. 
Note that a thorough investigation of the phase space of a particular model is not the goal of this paper. 
Instead, we investigate generic behaviors in community detection. 

\begin{table}[t!]
     \begin{center}
     \begin{tabular}{lrrcccccc}
     \toprule
      Dataset & $N = |V|$ & $L = |E|$ & $q^{\ast}_{\mathrm{mod}}$ & $q^{\ast}_{\mathrm{NBT}}$ \\ 
      \midrule
      karate club \cite{karateclub} & $34$ & $78$ & $1$ & $2$ \\
      dolphins \cite{dolphins} & $62$ & $159$ & $2$ & $2$ \\
      Les Miserables \cite{LesMiserables} & $77$ & $254$ & $2$ & $4$ \\
      football \cite{football} & $115$ & $613$ & $10$ & $10$ \\
      political books \cite{Newman2006politicalbooks} & $105$ & $441$ & $2$ & $3$ \\
      network science \cite{Newman2006PRE} & $379$ & $914$ & $4$ & $21$ \\
      C. elegans \cite{celegans} & $453$ & $2,025$ & $1$ & $5$ \\
      political blogs \cite{polblogs} & $1,222$ & $16,714$ & $3$ & $8$ \\
      protein \cite{konect:proteome,konect:2016:maayan-vidal,konect} & $2,738$ & $6,007$ & $1$ & $14$ \\
      power grid \cite{powergrid} & $4,941$ & $6,594$ & $10$ & $25$ \\
      Facebook ego \cite{FacebookEgoSNAP} & $4,039$ & $88,234$ & $15$ & $55$ \\
      chess \cite{chess,konect:2016:chess,konect} & $7,115$ & $55,779$ & $22$ & $45$ \\
      US airports \cite{Opsahl2010245} & $7,976$ & $15,677$ & $7$ & $17$ \\
      ca-HepTh \cite{caHepTh} & $8,638$ & $24,806$ & $33$ & $82$ \\
      Enron \cite{Enron1,Enron2} & $33,696$ & $180,811$ & $9$ & $93$ \\
      Epinions \cite{Epinions} & $75,888$ & $405,739$ & $6$ & $202$ 
      \\ \bottomrule
      \end{tabular}
      \caption{Estimated number of clusters of real-world networks using the spectra of the modularity matrix $q^{\ast}_{\mathrm{mod}}$ and non-backtracking matrix $q^{\ast}_{\mathrm{NBT}}$. Multi-edges, self-loops, and the direction and weights of edges are neglected in all networks.}
      \label{tableRealWorldNetworks}
      \end{center}
\end{table}

\begin{figure*}[ht!]
 \begin{center}
   \includegraphics[width=1.8 \columnwidth, bb=0 0 838 1204]{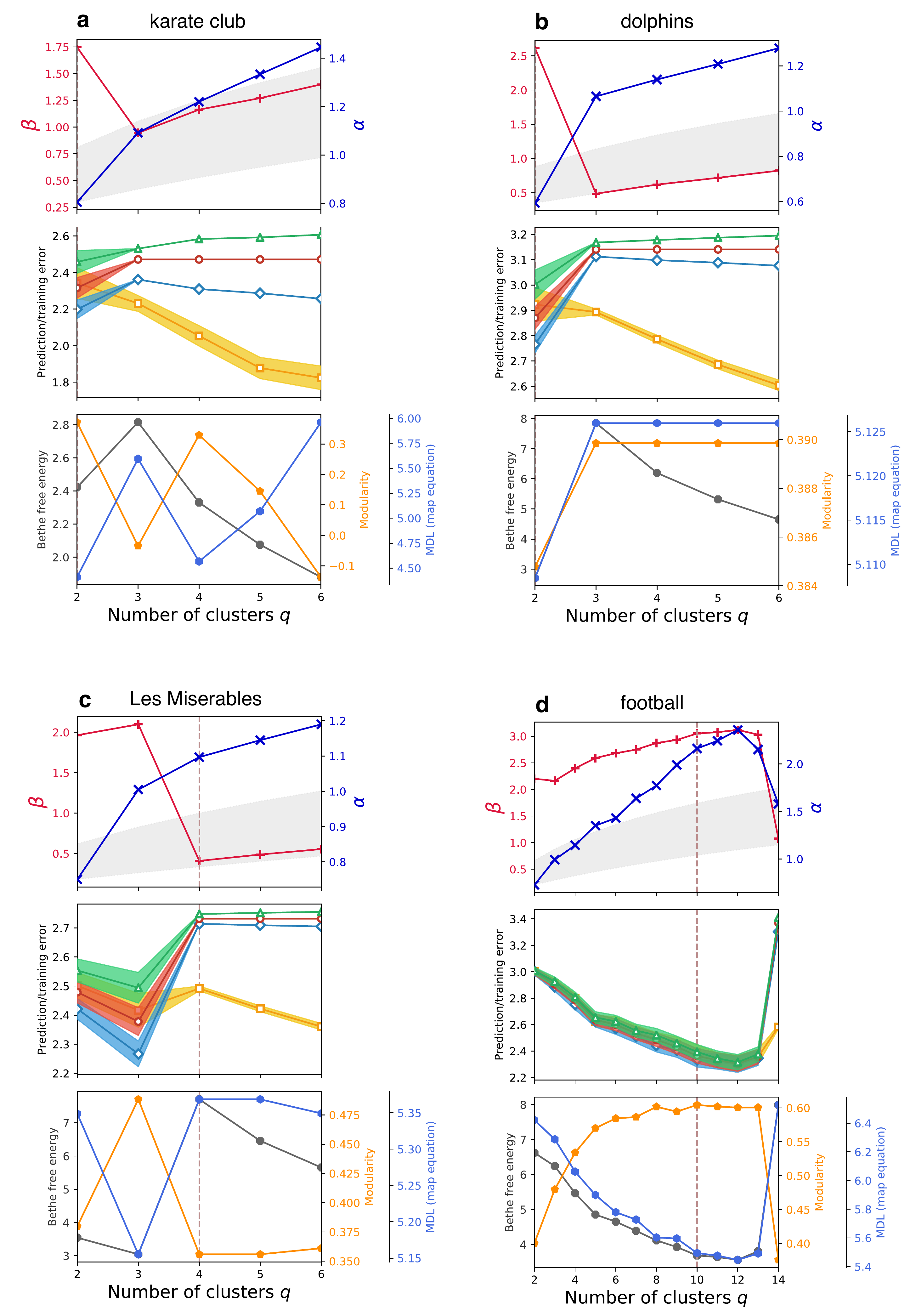}
 \end{center}
 \caption{Assessment of various criteria with respect to given inputs of $q$ for several real-world networks. 
 They are plotted in the same manner as in Figs.~\ref{LFR1Learned}--\ref{LFR6}. 
 The shaded parts in the cross-validation error plot indicate the standard errors. 
 }
 \label{RealWorldassessment1}
\end{figure*}

\begin{figure*}[ht!]
 \begin{center}
   \includegraphics[width=1.8 \columnwidth, bb=0 0 838 1204]{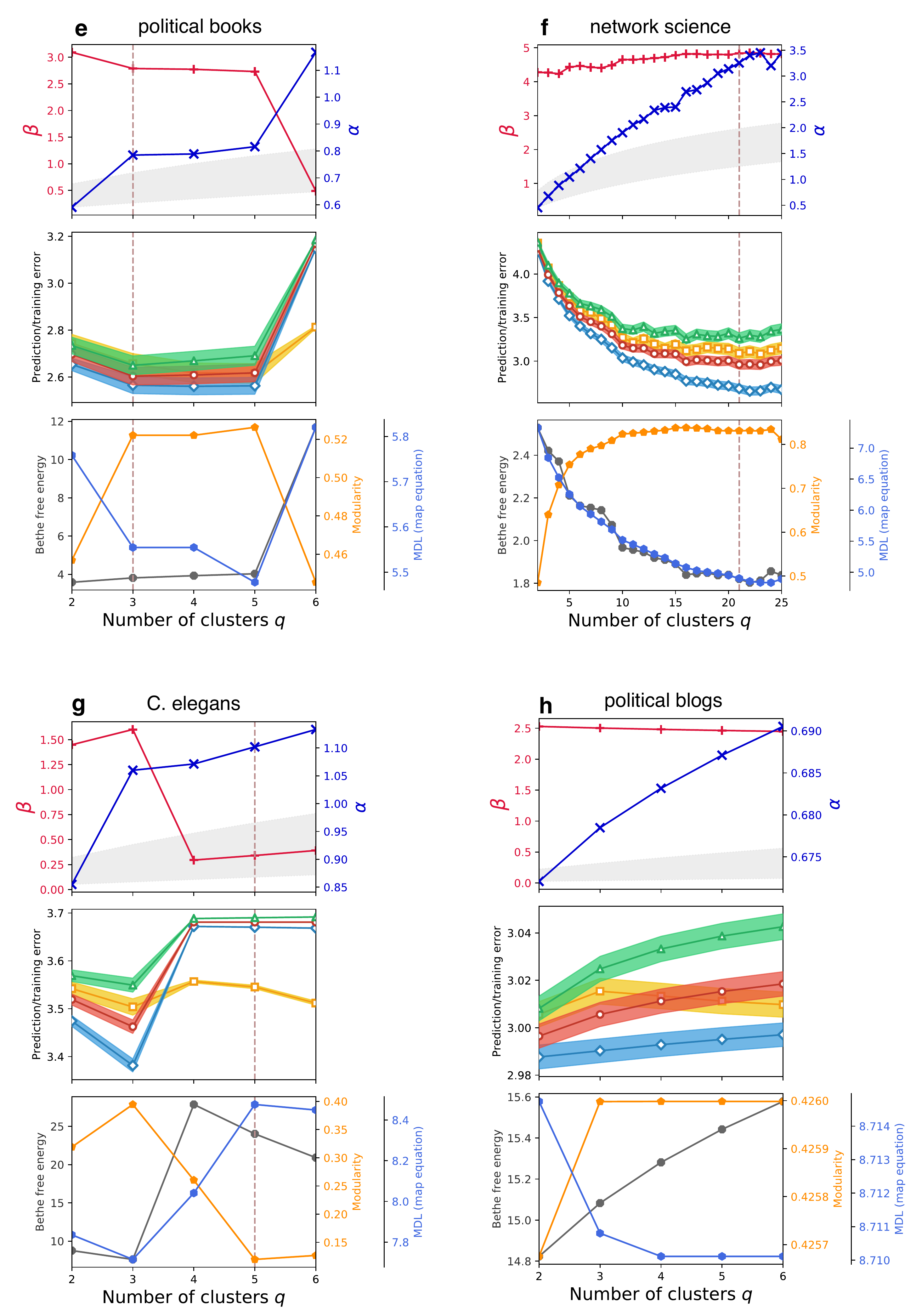}
 \end{center}
 \caption{Assessment of various criteria with respect to given inputs of $q$ for several real-world networks. 
 They are plotted in the same manner as in Figs.~\ref{LFR1Learned}--\ref{LFR6}. 
 The shaded parts in the cross-validation error plot indicate the standard errors. 
 }
 \label{RealWorldassessment2}
\end{figure*}

\begin{figure*}[ht!]
 \begin{center}
   \includegraphics[width=1.8 \columnwidth, bb=0 0 838 1204]{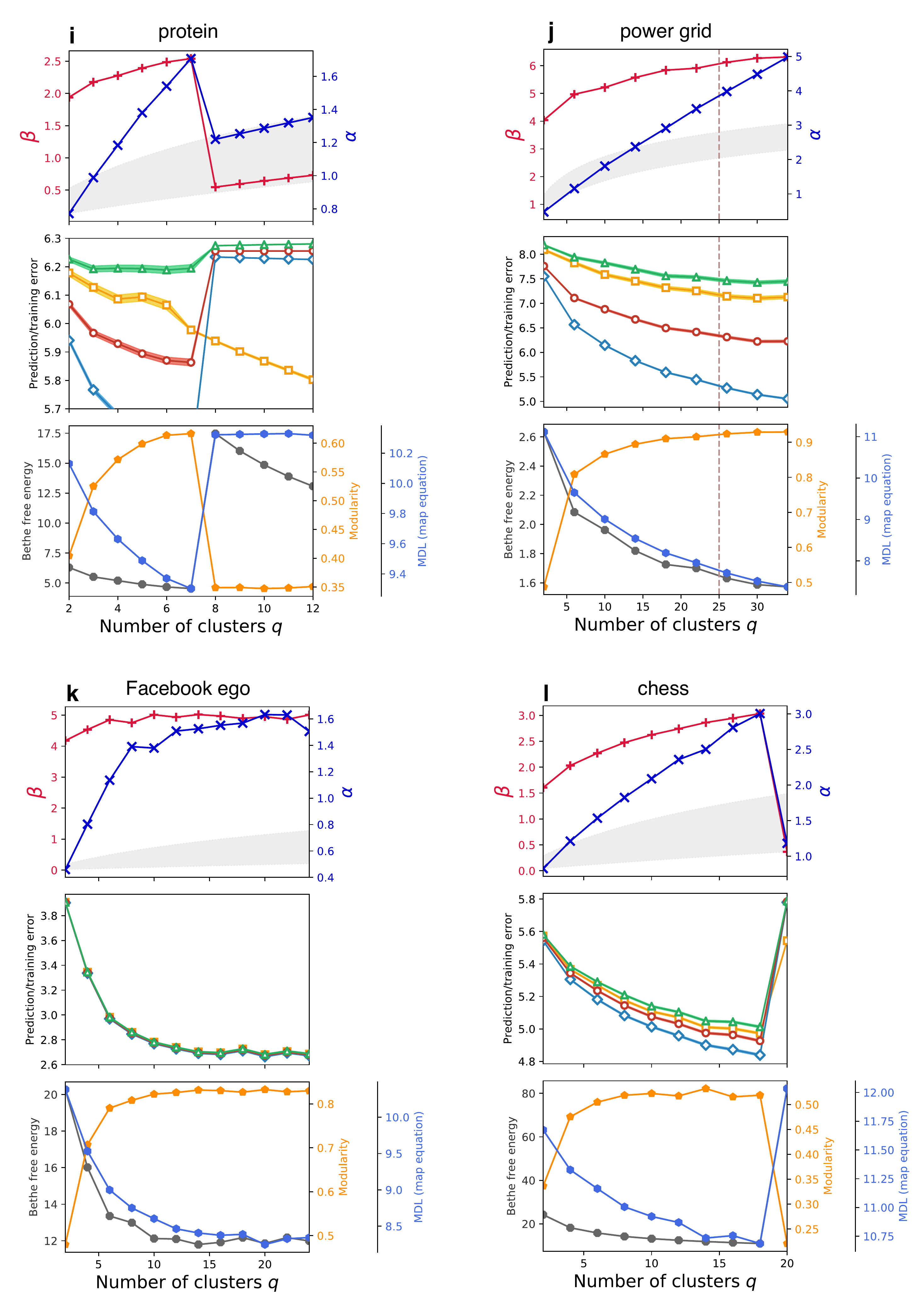}
 \end{center}
 \caption{Assessment of various criteria with respect to given inputs of $q$ for several real-world networks. 
 They are plotted in the same manner as in Figs.~\ref{LFR1Learned}--\ref{LFR6}. 
 The shaded parts in the cross-validation error plot indicate the standard errors. 
 }
 \label{RealWorldassessment3}
\end{figure*}
 
\begin{figure*}[ht!]
 \begin{center}
   \includegraphics[width=1.8 \columnwidth, bb=0 0 838 1204]{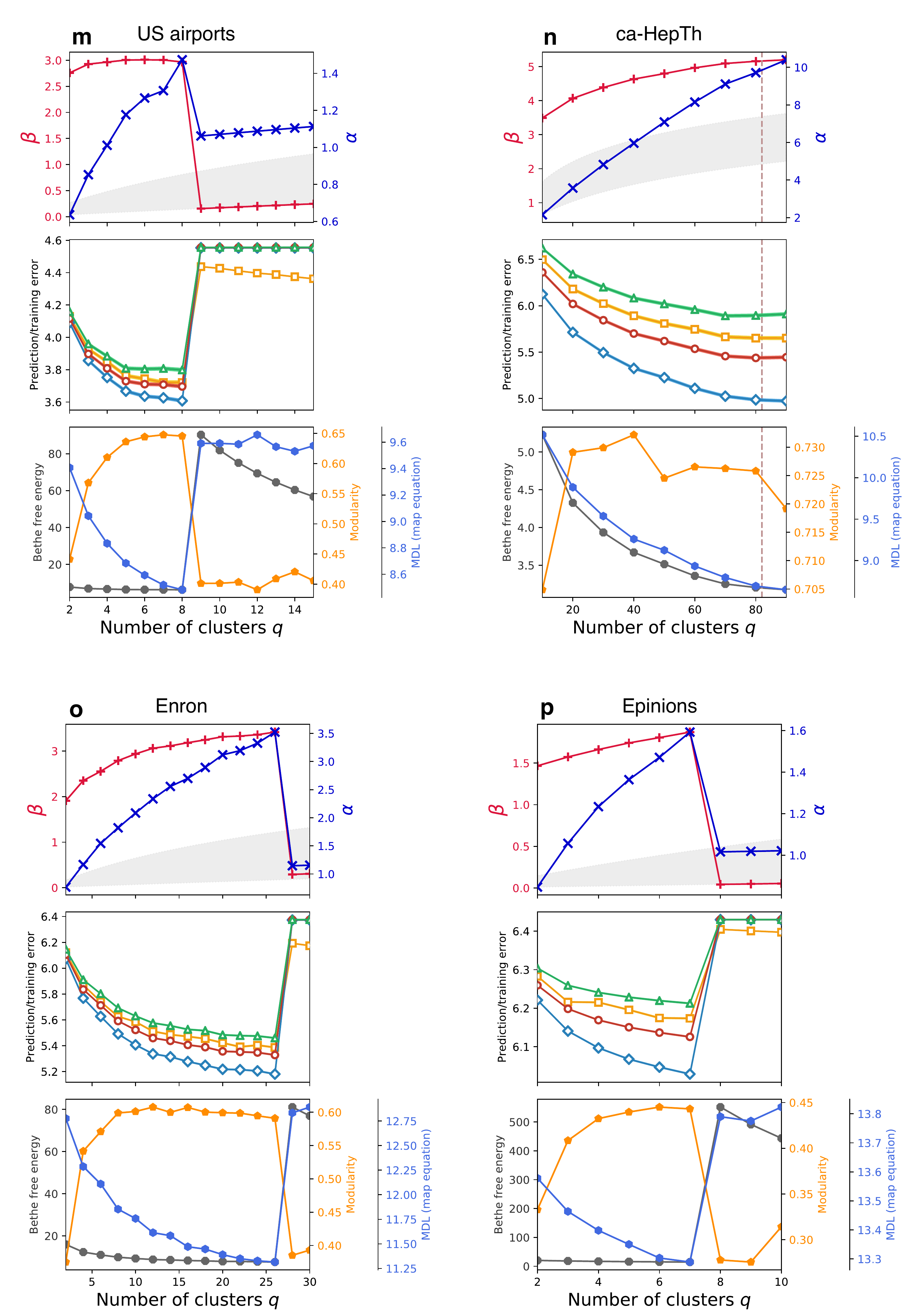}
 \end{center}
 \caption{Assessment of various criteria with respect to given inputs of $q$ for several real-world networks. 
 They are plotted in the same manner as in Figs.~\ref{LFR1Learned}--\ref{LFR6}. 
 The shaded parts in the cross-validation error plot indicate the standard errors. 
 }
 \label{RealWorldassessment4}
\end{figure*}

Let us next examine the performance of the assessment criteria on real-world networks. 
The basic information about each dataset is listed in Table~\ref{tableRealWorldNetworks} and the results of the assessment by each criterion are shown in Figs.~\ref{RealWorldassessment1}(a)--(d), \ref{RealWorldassessment2}(e)--(h), \ref{RealWorldassessment3}(i)--(l), and \ref{RealWorldassessment4}(m)--(p). 
For some networks, the inference algorithm does converge to the factorized state at some point as the input value of $q$ is increased; as far as we investigated, in many cases, the convergence to this trivial BP fixed point either supports a plausible value of $q^{\ast}$ or does not affect the assessment. 

It is known that the Bethe free energy tends to largely overfit for real-world networks \cite{Decelle2011a,KawamotoKabashimasbmBIX} when an affinity matrix of full degrees of freedom is used. 
However, with a restricted affinity matrix, the assessment by the Bethe free energy does not seem to be very wrong. 

Unlike the cases of synthetic networks, the behaviors of the assessment criteria are often very different from each other. 
For example, modularity tends to support a relatively small value for $q^{\ast}$, while the map equation tends to support a relatively large value. 
Assessment by the Bethe free energy and prediction errors can be close to either of them, and we cannot determine a similarity tendency. 
Note again that the inference algorithm does not optimize the minimum description length of the map equation; the partition is obtained such that the marginal likelihood is maximized and the minimum description length is measured only as a criterion for the assessment of the number of clusters. 
Another way to utilize the assessment by the minimum description length is to check whether the resolution limit \cite{Fortunato2007,KawamotoRosvall2015} affects the result. 
The estimates for the number of clusters $q^{\ast}$ by modularity and by the map equation can differ considerably when many small clusters exist, because their resolution limits are very different \cite{KawamotoRosvall2015}. 
When both modularity and the map equation support the same number of clusters $q^{\ast}$, it implies that the network is resolution limit-free \cite{FootnoteResolutionlimitfree}. 

\subsection{Assessment using the spectral methods}\label{Comparison:Spectral}
Finally, we examine the performance of the assessment of the number of clusters using the spectral methods that we explained in: Secs.~\ref{Algorithm:SpectralMethods} and \ref{Assessment:SpectralMethods}. 
The results of the estimates using the modularity matrix $q^{\ast}_{\mathrm{mod}}$ and non-backtracking matrix $q^{\ast}_{\mathrm{NBT}}$ are listed in Tables.~\ref{LFRtable} and \ref{tableRealWorldNetworks}. 
The estimates using the non-backtracking matrix are also shown in Figs.~\ref{LFR1Learned}--\ref{RealWorldassessment4} as dashed lines. 

Note that, because the leading eigenvalues can be quickly computed for sparse networks, the assessment using the spectral method can be conducted most easily. 
Overall, the assessment using the modularity matrix tends to underfit, while the assessment using the non-backtracking matrix tends to overfit, compared to the other criteria. 
Furthermore, for real-world networks, it is often the case that the spectral band of the modularity matrix cannot be clearly observed. 
Therefore, in many cases, it is also difficult to visually assess the number of clusters from the spectral density. 

The assessment using the non-backtracking matrix is often very accurate in the sense that it coincides with the planted value $q_{\mathrm{planted}}$ of an LFR network. 
The analysis with various values of $\mu$ was also analyzed in Ref.~\cite{Darst2014}. 
It is difficult to analyze what exactly causes overfitting in the cases of the real-world networks; one of the possibilities is that the community structure may not be the only structure that contributes to the eigenvalues outside of the spectral band, and those unknown structures cause overfitting when we focus on community detection.

\clearpage

\section{Assessment through visualization}\label{Visualization}
As we have observed, the validation curves of the criteria are often gradually saturated, particularly when the community structure is weak. 
In such a case, a criterion supports a range of values for $q^{\ast}$, instead of a single plausible value. 
Therefore, a finer inspection is required as a final step, if one wishes to determine the most plausible value of $q^{\ast}$. 

Visualizing how a network is partitioned for each input value of $q$ can be helpful for the assessment of the number of clusters. 
The alluvial diagram \cite{Rosvall2010} is a suitable tool for this purpose. 
It was originally introduced as a diagram to indicate the time evolution of a community structure. 
Here, we visualize the change in the partition for different values of $q$, rather than the change in the partition over time.
In the alluvial diagram, the results of community detection are aligned horizontally. For each partition, the set of vertices in the same cluster is expressed as a vertical bundle. Then, the same vertices in different partitions are smoothly connected. 
The alluvial diagram can be generated at Ref.~\cite{mapequationorg} using \texttt{.smap} files. 

The alluvial diagram also uses color tone to express the significance of the cluster assignment; the vertices with insignificant assignments are expressed by faint colors. 
We assess that the cluster assignment of vertex $i$ is not significant if $\max \psi^{i}_{\sigma}$ is less than a threshold. 
Here, we regard $\max \psi^{i}_{\sigma} > 0.9$ as a significant assignment.

\begin{figure}[t!]
 \begin{center}
   \includegraphics[width=0.9 \columnwidth, bb=0 0 582 1300]{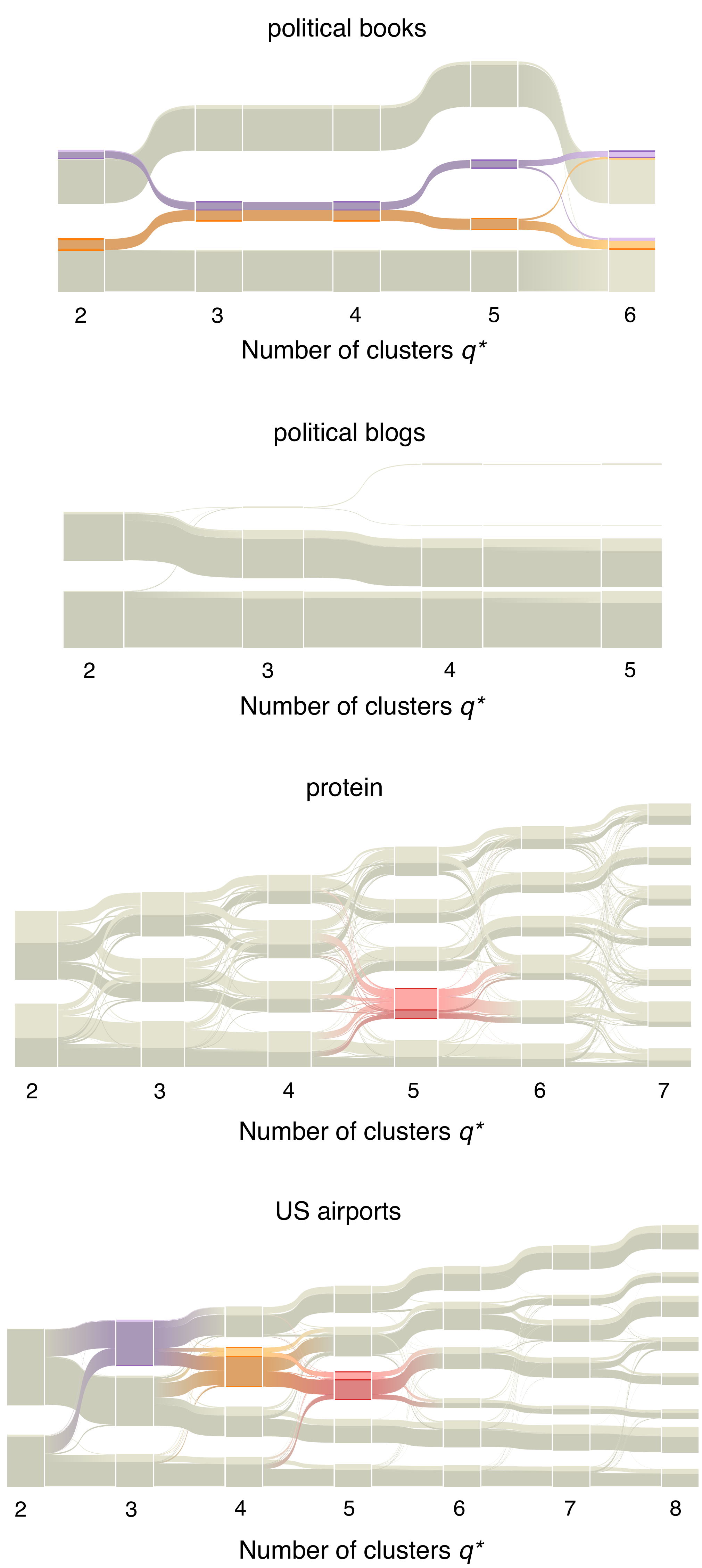}
 \end{center}
 \caption{Alluvial diagrams of the \textit{political books}, \textit{political blogs}, \textit{protein}, and \textit{US airports} networks. Some clusters are highlighted.}
 \label{AlluvialDiagrams}
\end{figure}

The alluvial diagrams of four real-world networks are shown in Fig.~\ref{AlluvialDiagrams}. 
As we can see from the \textit{political books} and \textit{political blogs} networks, the actual partition may be kept the same as we increase the input value of $q$. 
We can confirm that the partition with $q=3$ stably exists in the \textit{political books} network and the partition with $q=4$ is also consistent with the partitions with fewer clusters; i.e., it is a refinement of the partition with $q=3$, and the highlighted clusters in the middle belong to different clusters in the partition with $q=2$. 
For the \textit{political blogs} network, although modularity and the map equation support $q=3$ or $4$, in the end, we can confirm that the dominant structure does not change from the partition with $q=2$. 

In the case of the \textit{protein} network, for any choice of $q$, only a fraction of vertices belong to a specific cluster significantly. In other words, the network does not have a global community structure. 
Whereas the vertices with insignificant assignments exhibit a random-like behavior, the vertices with significant assignments roughly exhibit a hierarchical structure. 
According to the Gibbs prediction error $E_{\mathrm{Gibbs}}$ and its MAP estimate $E_{\mathrm{MAP}}$ in Fig.~\ref{RealWorldassessment3}(i), the partitions with $q=3$ or $4$ are supported. 
Although we cannot clearly see a qualitative difference in the alluvial diagram, if we look carefully, from the partition with $q \ge 5$, we can observe that a small fraction of vertices with significant assignments start to exhibit a non-hierarchical structure. 

The way the \textit{US airports} network is partitioned is also different from other networks. 
When we focus on the vertices with significant assignments, the resulting partitions do not constitute a hierarchical structure for small values of $q$, while they roughly do for large values of $q$. 
As various assessment criteria support the range of $5 \le q^{\ast} \le 8$, it seems to be plausible to select $q^{\ast}$ in this range.

\section{Summary and Discussion}\label{SummaryDiscussion}
We conducted a comparative analysis on the assessment of the number of clusters in community detection. 
Although we examined the performance of various algorithms and assessment criteria, an exhaustive analysis requires all possible combinations of the frameworks, algorithms, and assessment criteria. 
For example, an important missing part is the Monte Carlo method \cite{Nowicki2001, PeixotoPRE2014MonteCarlo,NewmanReinert2016}. 
The Monte Carlo method usually incorporates the prior probability distribution with respect to the affinity matrix $\ket{\omega}$; it plays the role of a penalty in the assessment of the number of clusters. 
Therefore, a comparative analysis including the Monte Carlo method would not only be a comparison of different algorithms, but also a comparison of different frameworks.  
In a broader sense, we should note that community detection also possesses some other fundamental issues as discussed in Ref.~\cite{Peel2017}. 

We confirmed that the assessment using the EM algorithm with BP and the corresponding prediction errors also provide plausible estimates in various synthetic and real-world networks, while the greedy algorithms tend to largely overfit. 
Note that it is not trivial that the BP algorithm performs reasonably for real-world networks, because the emergence of the so-called hard phase \cite{Decelle2011a} may deteriorate the performance when the planted number of clusters is large. 
Furthermore, the EM algorithm may be trapped in a local minimum depending on the choice of the initial condition of the model parameters. 

We also observed that the estimate of $q^{\ast}$ using the modularity matrix tends to underfit, while the estimate using the non-backtracking matrix tends to overfit. 
To the best of our knowledge, the assessment using the boundary of the spectral band of the modularity matrix has not been investigated in the literature. 


Finally, we proposed the utilization of the alluvial diagram for the assessment of $q^{\ast}$. 
Although there is the obvious issue that it is not applicable to the networks with a very large $q^{\ast}$, when it is not the case, the alluvial diagram is very useful, particularly when the network does not clearly have a global community structure. 

For the LFR networks and real-world networks, we do not know the number of clusters that are statistically significant. 
It may not coincide with the planted number of clusters or the number of clusters in the metadata. 
Therefore, we rely on the consistency among various criteria and algorithms for the plausibility of assessment. 
The tendency of overfit and underfit that we investigated in this paper represents the relative performance among the criteria and algorithms. 
Although there is no ground truth in a real-world network, estimating the number of clusters is a practical problem. 
At the end of the day, a practitioner selects a certain value (or a set of values) as $q^{\ast}$.

The code for the assessment using the modularity matrix is available at Ref.~\cite{Github-Modularity}. 
The code for the assessment using the other criteria in this paper, which can also generate \texttt{.smap} files, is available at Ref.~\cite{graphBIXurl}. 

\section*{acknowledgments}
T. K. thanks Jean-Gabriel Young for useful comments. 
This work was supported by Japan Society for the Promotion of Science KAKENHI Grants No. 26011023 (TK) and No. 25120013 (YK).

\bibliographystyle{apsrev}
\bibliography{bib-modBIX}

\begin{thebibliography}{79}
\bibitem[{Foo({\natexlab{a}})}]{FootnoteEMalgorithm}
\bibinfo{howpublished}{A formal derivation of the EM algorithm is the one using
  the variational expression, e.g., Ref.~\cite{KawamotoKabashimasbmBIX}.}
\bibitem[{Foo({\natexlab{b}})}]{FootnoteDetectabilityThreshold}
\bibinfo{howpublished}{Precisely speaking, the algorithmic detectability
  threshold that a certain algorithm fails to retrieve the planted structure is
  different from the information-theoretic detectability threshold that it is
  fundamentally impossible to retrieve the planted structure.}
\bibitem[{Foo({\natexlab{c}})}]{FootnoteFactorizedState}
\bibinfo{howpublished}{Of course, this can also be confirmed directly by
  observing the marginal distributions themselves.}
\bibitem[{Foo({\natexlab{d}})}]{FootnoteResolutionlimitfree}
\bibinfo{howpublished}{Although the word ``resolution limit-free'' is sometimes
  used for an algorithm, it seems more appropriate to say that a network is
  resolution limit-free for a given set of algorithms.}
\end{thebibliography}

\end{document}